\documentclass[a4paper,12pt]{article}
\usepackage{graphicx} 

\usepackage[top=2 cm, left=2 cm, right=2 cm, bottom=2 cm]{geometry}
\usepackage{subcaption}
\usepackage{xcolor}
\usepackage{amsmath}
\usepackage{float}
\usepackage{amsfonts}
\usepackage{mathrsfs}
\usepackage{mwe}
\usepackage{indentfirst}
\usepackage[colorlinks=true,
    linkcolor=blue,
    filecolor=magenta,      
    urlcolor=cyan]{hyperref}
\hypersetup{colorlinks=true,linkcolor=blue,urlcolor=blue}

\usepackage{bbold}
\usepackage{mathabx}

\title{\bf
Cartan Fluxes in $SU(3)$ Lattice Gauge Theory}

\author{
Tereza Mendes,$^a$ 
Luis E. Oxman,$^b$
Gustavo M. Simões$^a$
and 
Rafael C.S. Tonhon$^a$
\\[2mm]
$^a$ Instituto de Física de São Carlos,
     Universidade de São Paulo, IFSC--USP, \\[2mm]
     13566-590, São Carlos, SP, Brazil \\[2mm]
$^b$ Instituto de Física, Universidade Federal Fluminense, \\[2mm]
     24210-346 Niterói, RJ, Brasil
}

\begin{document}

\maketitle
\begin{abstract}
We propose and analyze a new method of detecting center vortices and monopoles
in lattice Yang-Mills theory. This procedure is sensitive to the intrinsic
degeneracy of the center charges, which play a crucial role in how these
topological objects interact and correlate with one another.
Our approach is based on fixing the Maximal Abelian gauge (MAG) and
decomposing the link configuration in a suitable way to look for
so-called Cartan fluxes, by projecting the gauge 
fields onto the Cartan subalgebra.
The method directly assigns the detection of monopoles to the 
roots of the gauge group, allowing a clearer and more robust 
characterization of the Abelian-charge content of the
gauge configuration.
Our discussion is general for $SU(N)$ gauge theory,
but we focus our applications on the $SU(3)$ case. For the $SU(2)$ case,
our proposed parametrization is equivalent to the 
standard one.
We present a numerical study of the monopole density in 
$SU(3)$ theory, obtained using our method. A sizable difference
in the number of monopoles is found between our proposed method and 
the standard one. 
Also, we observe a Weyl-symmetric distribution of monopole charges.
\end{abstract}

\section{Introduction}
Understanding confinement in 4D Yang-Mills theory requires studying the quantum vacuum, which is populated by topological structures such as percolating center-vortex worldsurfaces and monopole worldlines. Pioneering research established the importance of monopoles and percolating center vortices in this context, see e.g.\ \cite{tHooft:1981bkw,Mandelstam:1974pi,tHooft:1977nqb,tHooft:1979rtg,Mack:1978rq,Nielsen:1973cs}. As realized later, lattice simulations offer an ideal testing ground for such ideas, but the procedure for detecting topological excitations on the lattice must be chosen carefully. This typically involves a gauge-fixing step, intended to bring the thermalized gauge-link configurations as close as possible to ``Abelian-like'' field variables that emphasize some desired characteristics. Then, one implements a projection of the links onto a given subgroup of the gauge group, highlighting the interesting topological objects related to the above proposals.

Starting with center vortices in $SU(2)$ theory, one manner of detection
is by fixing the Direct Maximal Center (DMC) gauge with a subsequent 
projection onto $Z_2$, the center of $SU(2)$ 
\cite{DelDebbio:1998luz,Engelhardt:1999xw,Bertle:1999tw,Bertle:2000qv,Faber:1999gu,deForcrand:1999our}. 
Then, the center-projected plaquettes that yield a value $-\mathbb{1}$ are 
called $P$-plaquettes and are said to be pierced by a vortex guiding center
\cite{DelDebbio:1998luz}. Such plaquettes are, in fact, correlated with the 
location of the center-vortex guiding center in the unprojected link 
configuration \cite{Faber:1999gu}, making this a reliable vortex-finding
procedure. 
The technique led to the observation of \emph{Center Dominance} 
\cite{DelDebbio:1998luz,Bertle:2000qv}, as well as to the evaluation and 
study of several important quantities, such as vortex density 
\cite{DelDebbio:1998luz,Bertle:2000qv}, Pontryagin index 
\cite{Engelhardt:1999xw} and the vortex thickness 
\cite{Engelhardt:1998wu,Kovacs:2000sy}. 
Indeed, the latter property has been related to the mechanism behind the 
observed Casimir scaling of Wilson loops at intermediate distances 
\cite{Faber:1997rp}.
Also, this type of analysis provided a vortex-removal procedure, clarifying
the importance of center vortices for confinement \cite{deForcrand:1999our}.
Naturally, the same gauge condition can be extended to the case of $SU(3)$. 
In this case, the vortex density has a well-defined continuum limit 
\cite{Langfeld:2003ev} and center dominance is still observed, albeit more
subtly \cite{Trewartha:2015ida}. More recently, the detection of 
center-vortex worldsurfaces was addressed in the DMC gauge with the use of
3D visualization techniques \cite{Biddle:2019gke}.
This revealed, at fixed time slices, a network formed by center-vortex loops
and a primary cluster of center-vortex lines that can meet in groups of three. 
These matching points, which reflect the conservation of center charge modulo
$3$, are known as {\em center monopoles} \cite{Spengler:2018dxt}. The center-vortex network was also recently studied in
the $N=4$ case \cite{LeinweberOxman2025}.

Another type of monopole-like object was extensively analyzed following a 
different procedure. In this setting, after
fixing the Maximal Abelian Gauge (MAG), the link variables are projected onto
the Cartan subgroup, i.e.\ the diagonal subgroup of $SU(N)$, 
with the goal of identifying suitable monopole currents.
This was implemented for the $SU(2)$ case in Refs.~
\cite{Suzuki:1989gp,Ivanenko:1990xu,DelDebbio:1991fp,Hioki:1991ai,Trottier:1993nu,Shiba:1994ab,Stack:1994wm,Bali:1996dm,Ambjorn:1999ym}.  
The resulting Abelian-projected field has a residual $U(1)$ gauge symmetry, 
based on which the algorithm proposed (for compact QED on the lattice) by 
DeGrand and Toussaint \cite{DeGrand:1980eq} can be applied. 
Important concepts such as {\em Abelian Dominance} emerged from this method
of detection of monopoles.
Other steps taken in these studies were measurements of monopole density 
and the falsifiability of the monopole-gas scenario of confinement, which
are key to the current understanding of confinement. 
Similar investigations were carried out for $SU(3)$ in
\cite{Stack:2002ysv,Stack:2002kh,Tucker:2002qu,Sakumichi:2014xpa,Suganuma:2018rvq,Bonati:2013bga}, 
considering different procedures for the MAG implementation and for the
Abelian projection of the gauge links. Moreover, these studies differ in
the prescription for monopole detection, i.e.\ the procedure for defining 
Abelian field tensors associated with the residual $U(1)^2$ symmetry in the
$SU(3)$ case, as well as the extension of the DeGrand-Toussaint 
algorithm for obtaining monopole currents.
In Refs.~\cite{Stack:2002ysv,Stack:2002kh,Tucker:2002qu}, the $U(1)$-detection 
prescription was applied independently to the three diagonal entries of the
projected link variables, obtained by determining the diagonal $SU(3)$
element closest to the link variable. Also, a modified version of the MAG was 
introduced.
In Refs.~\cite{Sakumichi:2014xpa,Suganuma:2018rvq}, on the other hand, 
the projection was done by taking explicitly into account the Cartan subalgebra
components of the diagonal link variable.
In particular, these projected variables have effectively captured the
confining properties of heavy quarks in the fundamental representation with
remarkable accuracy \cite{Sakumichi:2014xpa}. 
For detecting monopoles, however, the diagonal entries of the projected link
were taken independently in \cite{Suganuma:2018rvq}.
An alternative approach was followed in Ref.~\cite{Bonati:2013bga} for 
studying monopoles at finite temperature, with a different 
parametrization of the projected links, yielding two 
unconstrained phases to which a modified DeGrand-Toussaint algorithm 
was applied. 
This led to the identification of two types of thermal monopoles, and the 
authors studied their attractive/repulsive interaction as well as their 
condensation above and near the critical temperature.

The MAG can also be used to study center vortices in what is called Indirect
Maximal Center (IMC) gauge 
\cite{Ambjorn:1999ym,DelDebbio:1996lih,Langfeld:1997jx,DelDebbio:1997ke,Alexandrou:1999vx,deForcrand:2000pg}, 
which is useful for probing the correlation of center vortices with monopoles.
In fact, in Ref.~\cite{Ambjorn:1999ym}, the authors observed that, 
in $SU(2)$, 97\% of the elementary cubes containing a monopole have at least
a pair of $P$-plaquettes, showing a correlation between these objects.
Moreover, the analysis of the flux distribution around the faces of the cube
containing the monopoles reveals that the flux is concentrated precisely on
this pair, a feature referred to as {\em flux collimation}.
A (less strong) correlation between center vortices and 
monopoles was also observed in the $SU(3)$ case, by using similar techniques
\cite{Stack:2002ysv,Stack:2002kh,Tucker:2002qu}
and the generalized definition of the MAG mentioned above.

In general, the detection of $SU(N)$ vortices on the lattice takes into 
account only their center charges, given by the number $k=1,\dots,N-1$ 
in the contribution $e^{2\pi i\, k/N}$ to a Wilson loop that is linked
once by a center vortex. (See \cite{Greensite} for a review.)
However, a numerical investigation fully considering the positions of the 
vortex guiding centers and how the flux of these objects is distributed
is still lacking. 
In particular, the exponent of the projected plaquette variables, which in
general takes values in the Cartan subalgebra ---and is thus called 
{\em Cartan flux}--- is needed to investigate if the above-mentioned property 
of flux collimation, observed for $SU(2)$, is also verified in the $SU(3)$ case.
Moreover, an active line of research
\cite{Oxman:2018dzp,Oxman:2018nqe,Junior:2019fty,Junior:2022bol,Junior:2023tjg,Junior:2024urr}
proposes these degrees of freedom as essential to relate the center-vortex 
ensemble with confinement. 
This proposal is based on parametrizing a condensate of elementary center 
vortices and monopoles attached to them, forming chains \cite{Oxman:2018dzp,Oxman:2018nqe}. The elementary 
vortices, for example, can meet at a single point if there are $N$ of 
them, each carrying a different weight, out of the $N$ possible values 
associated with $k=1$. 
Another example would be a {\em nonoriented center vortex}, where a pair of collimated fluxes is attached to a monopole, carrying different weights as they enter and leave, with the monopole accounting for the difference between the fluxes on either side. This delicate matching is indeed invisible to the traditional ways of detecting these objects, motivating 
the search for a new method of measuring Cartan fluxes on the lattice.

In this work, we propose such a method to characterize $SU(3)$ Cartan fluxes,
in a way that naturally assigns to vortices and monopoles their appropriate
charges.
Here, we mainly focus on monopole detection and the procedure for mapping out
the Cartan fluxes on the lattice. 
The first step in our approach is to fix the gauge according to the usual MAG
definition in a Weyl-symmetric way, so as not to introduce any bias in
the detection of Cartan fluxes and monopole charges.  
We then project the gauge-fixed link variables onto Cartan-basis fields, in
a manner similar to what was done in 
\cite{Sakumichi:2014xpa,Suganuma:2018rvq}, but without the maximization
procedure. In the third step, we parametrize the Abelian-projected plaquettes
using the basis of the Cartan subalgebra. 
In particular, we use the intrinsic periodicity given by
the algebra's simple roots to define a non-Abelian analogue for the 
DeGrand-Toussaint algorithm mentioned above.
This Cartan-based procedure is to be contrasted with the
previously employed ones in Refs.~\cite{Sakumichi:2014xpa,Suganuma:2018rvq,Stack:2002ysv,Stack:2002kh,Tucker:2002qu}. 
Finally, we describe our method in detail and 
show first results from our simulations.

\vskip 3mm
We begin by reviewing the concepts of topological observables for QED on the lattice and their extension to the $SU(2)$-theory case in Section \ref{corresu2}. Then, we discuss center vortices and Cartan fluxes
for $SU(N)$, as well as previous algorithms for monopole detection.
Next, in Section \ref{corrsuN}, we outline our proposed projection method
in the general case.
Details of our parametrization are given in Appendices \ref{Cartan-basis} 
and \ref{App:weights}.
Then, we discuss our implementation of the MAG on the lattice in Section
\ref{MAG}. The numerical results and our conclusions are presented in Sections \ref{results} and \ref{conclusions}, respectively.


\section{Topological Objects on the Lattice}
\label{corresu2}

\subsection{Monopole Detection in Compact QED}

\vskip 3mm

In compact lattice QED, the physical content associated with the 
link variable $U_\mu (x) = e^{i\theta_\mu(x)}$ can be studied by means 
of the field strength tensor\footnote{The lattice derivative is 
calculated using the forward prescription 
$\partial_\mu f(x) = f(x+\hat{\mu})-f(x)$.}
\begin{equation}
    f_{\mu\nu}(x) = \partial_\mu \theta_\nu(x)-\partial_\nu \theta_\mu(x)\,,
\qquad \mu,\nu=1,\dots,4\;.
    \label{theta}
\end{equation}
Here we consider
$\theta_\mu(x)$ in $\left[-\pi,\pi\right]$, so $f_{\mu\nu}(x)$
may take values in the range $\left[-4\pi,4\pi\right]$.
Although the total flux of the dual field strength $\widetilde{f}_{\mu\nu}=\frac{1}{2}\varepsilon_{\mu\nu\rho\sigma}f_{\rho\sigma}$ through the faces of an 
elementary cube
is $0$, monopoles can be detected following the 
method introduced by DeGrand and Toussaint  
\cite{DeGrand:1980eq}.
In this procedure, one looks for the decomposition 
\begin{equation}
\label{HDecomposition}
  f_{\mu\nu}\,=\, \bar{f}_{\mu\nu} + 2\pi\, n_{\mu\nu}\,,\qquad
 \bar{f}_{\mu\nu}\in\left[-\pi,\pi\right]\,, \qquad
 n_{\mu \nu} \in \mathbb{Z}\;.
\end{equation}
That is, whatever the value of $f_{\mu\nu}$, a multiple of $2\pi$ is subtracted
to obtain $\bar{f}_{\mu\nu}$ in the specified range, and such multiple defines
$n_{\mu\nu}$.\footnote{When monopoles are present, $\bar{f}_{\mu \nu}$ is not
a strength tensor obtained from a vector potential.} To obtain the total
monopole four-current $j_\mu$ in a unit cube centered at
$\tilde{x}$ defined by coordinates ${\tilde{x}}_\mu=x_\mu+0.5$, 
we use the formula \cite{Suganuma:2018rvq}
\begin{equation}
j_\mu(\tilde{x}) \,=\,
- \partial_\nu \widetilde{n}_{\mu \nu} \,, \qquad  
\widetilde{n}_{\mu \nu} \,=\, \frac{1}{2}\, \epsilon_{\mu\nu\rho\sigma}
\, n_{\rho \sigma} \;,
\label{mmu}
\end{equation}
where 
$\varepsilon_{\mu\nu\rho\sigma}$ is the
totally antisymmetric Levi-Civita tensor, there is an implicit sum over $\nu$, $\rho$, and $\sigma$, and all indices run from $1$ to $4$.
Since the total flux of $\widetilde{f}_{\mu\nu}$ over any cube vanishes (by the Bianchi identity), $j_\mu$ could be
evaluated by the exact same expression with $\bar{f}_{\mu\nu}$  instead of $n_{\mu\nu}$ and a global minus sign.

\begin{figure}[H]
    \centering
    \includegraphics[width=0.4\linewidth]{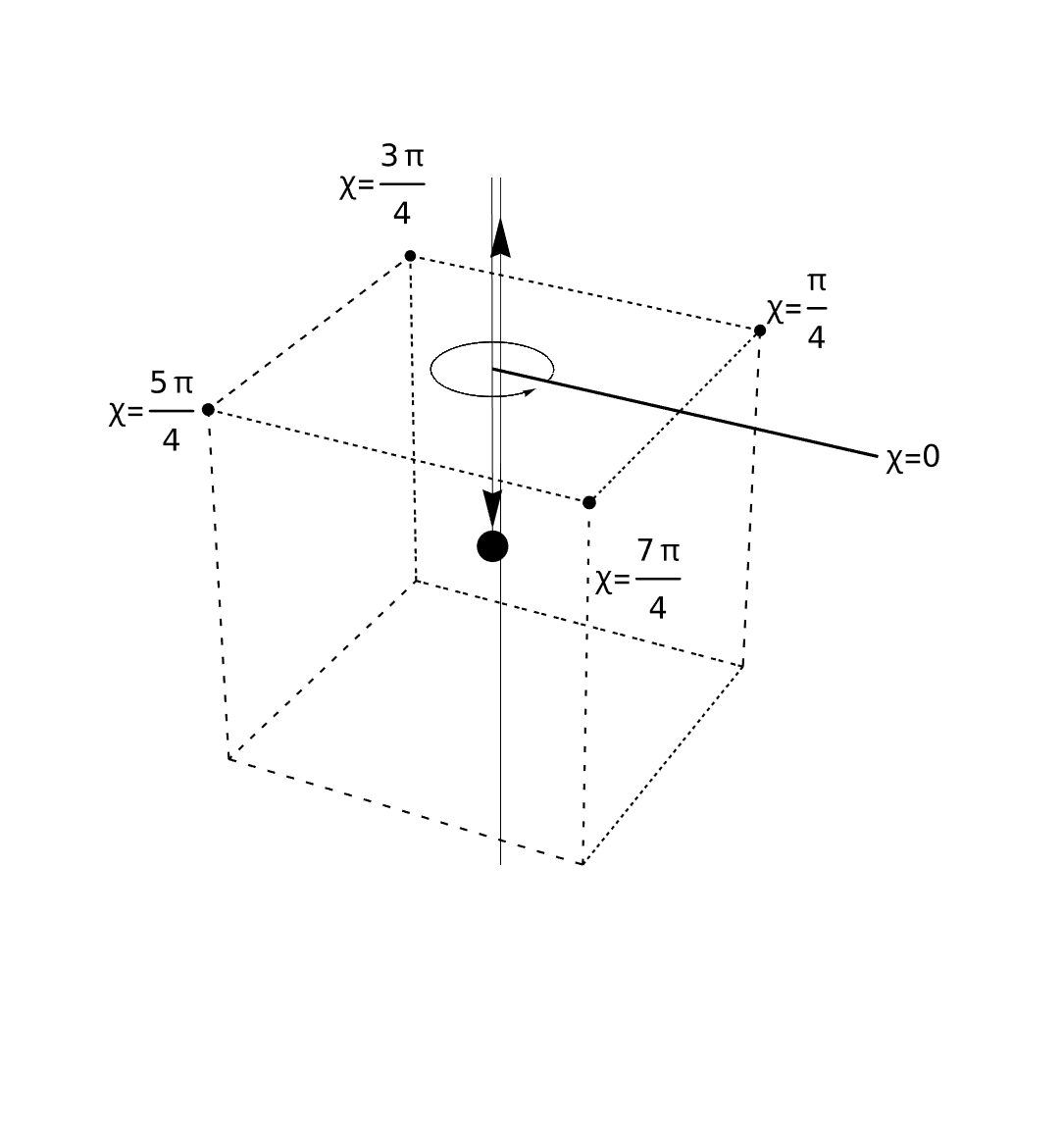}
    \caption{The downward Dirac string attached to a monopole can be moved, via
    a gauge transformation, to a physically equivalent upward semi-infinite
    string. }
    \label{MovingDiracString}
\end{figure}

The configuration $2\pi n_{\mu\nu}(x)$ is the lattice representation
of Dirac strings: a  tensor field $2\pi n_{\mu \nu}$, $n_{\mu \nu} \in \mathbb{Z}$, has
no effect on Wilson loops or the Wilson action.
Indeed, the plaquettes with $n_{\mu \nu} \neq 0$ can be changed by a gauge 
transformation, but the locations of elementary cubes with nontrivial total
flux of $\widetilde{n}_{\mu \nu}$ (monopole locations) cannot.
 For example, consider a given time slice and suppose a 
field $n_{ij}$ ($i,j=1,2,3$) that is nontrivial on the plaquettes pierced by
a vertical line running downward from $x_3=+\infty$,
passing through sites of the dual lattice and ending at a monopole.
This is illustrated in Fig.\,\ref{MovingDiracString}, in the vicinity
of the monopole. 
We see that $n_{12}$ equals $-1$ on the top horizontal plaquette. 
Now, we can perform a gauge transformation 
$\theta_\mu(x) \to \theta_\mu(x) + \chi(x+\hat{\mu}) - \chi(x)$,
where $\chi$ is the polar angle with respect to the vertical 
axis\footnote{Here, the polar angle $\chi$ is defined with a cut that does 
not contain any lattice site.}, which adds $2\pi$ to $f_{ij}$ at the 
plaquettes pierced by this axis, thus increasing the horizontal 
plaquettes $n_{12}$ by $1$ everywhere.
This has the net effect of erasing the original string and creating a new 
one running upwards from $x_3=-\infty$ up to the monopole, while leaving
$\bar{f}_{ij}$ invariant.

\subsection{Detection Tools in Yang-Mills Theory}
\label{current-d}

As discussed in the Introduction, a large amount of information has been 
gained over the years about center vortices and monopoles on the lattice. 
The main developments can be classified into vortex-only and monopole-only
scenarios, as well as studies of their correlations.

The detection of center vortices in Yang-Mills theory, in both the direct and
indirect maximal center gauges, is based on projecting gauge-fixed link 
variables onto the center of the gauge group.\footnote{The center of a group
$G$ is the subgroup formed by the elements that commute with all elements of
$G$.} In the case of $SU(2)$, the center is the $Z(2)$ subgroup 
$\{\mathbb{1}, -\mathbb{1}\}$. In this manner, plaquette variables computed
from the projected link variables allow for the vortex detection: those 
yielding the nontrivial element $-\mathbb{1}$ are identified as plaquettes 
pierced by a center vortex (guiding center).

When it comes to monopoles, they were detected by identifying effective 
Abelian degrees of freedom together with a prescription similar to that 
developed by DeGrand and Toussaint, described in the 
previous subsection. 
More precisely, one starts by fixing the MAG via the maximization of an 
appropriate functional (see Section \ref{MAG}).
In this gauge, the link variables are brought as close as possible to 
diagonal elements, i.e.\ to gauge-equivalent $SU(N)$ elements with
minimal off-diagonal components. In particular, the prescription is
unambiguous in the $SU(2)$ case, since there is only one diagonal generator,
given by the Pauli matrix $\sigma_3$.
Then, for each $U_\mu(x)\in SU(2)$, the Abelian-projected link 
variable $C_\mu(x)\in SU(2)$ is obtained as the closest diagonal group
element with respect to the distance 
Tr$\left[\left(U_\mu-C_\mu\right)\left(U_\mu-C_\mu\right)^\dagger \right]$,
i.e.\ by maximizing the overlap 
${\rm Re}{\rm Tr}\left(U_\mu C^\dagger_\mu\right)$, where
${\rm Tr}$ is the trace in color space. This defines the
Abelian-projected link variable
\begin{equation}
C_\mu(x) \,=\, e^{i \theta_\mu \sigma_3} \,=\,\left(\begin{array}{ccc}
         e^{i\theta_\mu(x)}& 0   \\
         0 & e^{-i\theta_\mu(x)}  
    \end{array}\right) \;,
    \label{Csu2}
\end{equation}
which can be used to calculate the Abelian-projected plaquettes
$\,C_{\mu\nu}\equiv e^{if_{\mu\nu}\sigma_3}$, 
where $f_{\mu\nu}$ is given by Eq.\,\eqref{theta}. 
Following this procedure, monopoles in $SU(2)$ were detected by simply 
applying Eq.\,\eqref{HDecomposition} to the phase $\theta_\mu$ in 
Eq.\,\eqref{Csu2}. In this case, the Dirac string for an elementary monopole
carries a flux $\,2\pi \,n_{\mu \nu}\, \sigma_3$, where $n_{\mu \nu} = \pm 1$
and the sign determines the orientation of the string. 
This flux contributes trivially to the (Abelian-projected)
Wilson loop and action, since $e^{\pm i2\pi\sigma_3 }=\mathbb{1}$.

As mentioned in the Introduction, it is interesting to note
that a strong correlation between monopoles and vortices in $SU(2)$ was 
observed on the lattice \cite{Ambjorn:1999ym,dF-P}. In the first 
reference, relying on the IMC gauge, it was found that 97\% of the detected
monopoles sit on top of vortices. It was also established that about 61\% of
the vortex lines have no monopoles, 31\% contain a monopole-antimonopole pair,
and 8\% of closed vortex lines have an even number of pairs, with monopoles
alternating with antimonopoles.
Besides these percentages, which refer to correlations with the vortex guiding
centers, correlations were also observed at the level of the full unprojected
lattice. This was done by studying the action density distribution on the faces
of the cube surrounding an elementary monopole. This distribution was found to
be highly asymmetric, with almost all action deviations from the mean 
concentrated on the P-plaquettes. This asymmetry could also manifest itself in
the flux distribution associated with the Abelian-projected variables, when 
analyzing the flux $\bar{f}_{\mu \nu}\sigma_3$ through the plaquette $(x,\mu,\nu)$ of a surface
surrounding an elementary monopole. Instead of a uniform flux distribution, it
is reasonable to expect a pair of fluxes leaving the monopole, collimated on
the typical scale of the center-vortex thickness and each carrying a flux
$\pi\sigma_3$. In particular, when computing Abelian-projected Wilson loops,
encircled collimated objects would contribute a nontrivial center element 
$e^{\pm \pi\sigma_3} = -\mathbb{1}$. Here, the sign in the exponent depends on
the relative orientation of the vortex and the loop. As discussed in 
Ref.~\cite{Ambjorn:1999ym}, the collimation was instead explored by studying
doubly charged Polyakov lines. The results showed that the Abelian-projected
lattice is not described by a monopole Coulomb gas, but by a collimated 
center-vortex structure.

\vskip 3mm
With regard to $SU(3)$, the center is the $Z(3)$ subgroup 
$\{\mathbb{1},e^{i2\pi/3}\mathbb{1},e^{i4\pi/3}\mathbb{1}\}$. This time, 
there are two nontrivial center elements. As $e^{i4\pi/3} = e^{-i2\pi/3}$, 
a center-vortex line characterized by a center charge $k=2$ can be thought
of as a unit-charge center vortex with opposite orientation.
Since the Cartan subgroup is two-dimensional, 
monopole detection is no longer clearly defined, and
a variety of methods can be found in the literature.
These methods may differ in one or more of the following three steps:
(i) the MAG gauge-fixing functional, (ii) the Abelian projection, and
(iii) the generalized DeGrand-Toussaint procedure. 

In Ref.~\cite{Stack:2002ysv}, prescriptions closely following the ones
in $U(1)$ and $SU(2)$ were considered for the MAG functional
and for the generalization of the
DeGrand-Toussaint algorithm 
(see Section \ref{Cartanbased}.)
In that work, center vortices were also located using the IMC gauge.
The fraction of times that $n$ faces of the cube dual to a magnetic-current
link are pierced by a $Z(3)$ flux was observed and the obtained values at 
$\beta =6$ were $0.16(1)$ ($n=0$), $0.74(1)$ ($n=2$), $0.07(2)$ ($n=3$), 
$0.02(2)$ ($n=4$), $0.001(1)$ ($n=5$), and $0.001(1)$ ($n=6$).  
On the other hand, no studies of the action-density distribution using 
unprojected variables, nor of the flux distribution of Abelian-projected
fluxes around monopoles, were performed.

In fact, in order to study the Abelian-projected lattice
for $SU(N)$ with $N \geq 3$, it is essential to understand which fluxes
generalize the $\pm \pi \sigma_3$ and $\pm 2\pi \sigma_3$ fluxes 
respectively carried by a collimated center vortex and a monopole in $SU(2)$.
Moreover, a consistent three-step procedure should be implemented to properly
detect monopoles and to map the fluxes carried along the thick collimated
vortices, guided by the thin-center-vortex network detected via the IMC gauge.
The objective of the next two sections is to fill this gap in the currently 
available detection tools, thereby opening the possibility of a deeper
characterization of the confining Yang-Mills vacuum.

\section{Cartan-Based Procedure}
\label{corrsuN}

\subsection{The Fluxes of Collimated Center Vortices}
\label{f-col}

Studies on the Abelian-projected lattice are based on bringing the link 
variables $U_\mu \in SU(N)$ to the MAG, followed by a projection onto Cartan
(diagonal) variables $C_\mu(x)$. As the Cartan subalgebra is
$(N-1)$-dimensional, we have
\begin{equation}
\label{algebraicparametrization}
    C_\mu(x)\,=\,e^{i\theta_\mu(x)\,\cdot\, T}\,,\qquad
\theta_\mu(x) \cdot\, T \,\equiv\; \theta_\mu(x)|_q\, T_q \;,
\end{equation}
where the repeated index is summed over $q=1, \dots, N-1$ and $T_q$ are the Hermitian basis elements defined in Eq. \eqref{dia}.
The corresponding projected plaquette variables
\begin{eqnarray}
    C_{\mu\nu}(x) \,\equiv\, C_\mu(x)\,C_{\nu}(x+\hat{\mu})\,
C^\dagger_{\mu}(x+\hat{\nu})\,C^\dagger_{\nu}(x) 
\end{eqnarray}
are then given by
\begin{equation}
\label{algebraicplaquette}
  C_{\mu\nu}(x) \,=\,  e^{i f_{\mu \nu} \,\cdot\, T}\,,\qquad
    f_{\mu \nu} \,\cdot\, T \,\equiv\, f_{\mu \nu}\vert_q\, T_q\,,\qquad
    f_{\mu \nu} \,=\, \partial_\mu \theta_\nu - \partial_\nu \theta_\mu\;.
\end{equation}
Naturally, $f_{\mu\nu}\cdot T$ is the Cartan flux in this case and equates to the magnetic flux through the plaquette $(x,\mu,\nu)$. Notice that $\theta_\mu$ and $f_{\mu \nu}$ are $(N-1)$-tuples, with
the components $f_{\mu \nu}|_q$ defined from the respective
components $\theta_\mu|_q$ as in Eq.\,\eqref{theta}.
Notice also that the relation between the (diagonal) elements of $C_\mu(x)$ and the ``angles'' $\theta_\mu(x)$ is not as direct as it was for
the $SU(2)$ case.

\vskip 3mm
Center vortices were detected by using the remaining $U(1)^{N-1}$ gauge 
symmetry to drive $C_\mu$ as close as possible to a $Z_\mu$ configuration in 
the center of $SU(N)$, formed by the elements 
\begin{equation}
e^{i\frac{2\pi k}{N} }\,\mathbb{1}\,,\qquad k=0, \dots, N-1 \;,
\end{equation}
and looking for nontrivial plaquettes
\begin{equation}
    Z_{\mu\nu}(x) \;=\, Z_\mu(x)\,Z_{\nu}(x+\hat{\mu})\,
    Z^\dagger_{\mu}(x+\hat{\nu})\,Z^\dagger_{\nu}(x) \;.
\end{equation}
By construction, these vortices are thin. They are expected to guide the flux distribution of thick collimated objects existing before the center projection. In this respect, it is interesting to note that the $Z(N)$ fluxes do not entirely capture the complexity of collimated fluxes. For instance, consider an Abelian-projected configuration such that
\begin{equation}
f_{\mu \nu} \cdot T \,=\, a_{\mu \nu}\, \beta_i \cdot T \,,\qquad   
\beta_i \,=\, 2N\, \omega_i \;,
\label{betadef}
\end{equation}
where $\omega_i$, $i = 1, \dots , N$, is a given weight of the defining 
representation (see Appendix \ref{App:weights}). Here, $\beta_i$ is an 
$(N-1)$-tuple, while the variables $a_{\mu \nu}$ are real numbers.
The Wilson loop for a curve $\mathcal{C}$ on the Abelian-projected lattice is
\begin{equation}
    \mathcal{W}_\mathcal{C} \,=\, e^{i\, \sum_{S} f_{\mu \nu} \cdot\, T} \;,
\end{equation}
where $\sum_S = \sum_{(\mu \nu) \in S}$ denotes the sum over the plaquettes
that form any surface $S$ whose border is $\mathcal{C}$. Now, consider a 
collimated distribution such that $\,\sum_{S} a_{\mu \nu} \approx \pm 2\pi$, 
where the sign determines how the flux is oriented in the 4D discretized 
spacetime. In this case, the encircled total flux and the Wilson loop are, 
respectively,
\begin{equation}
\sum_{(\mu \nu)} f_{\mu \nu} \,\approx\, \pm 2\pi \beta_i \cdot T 
\,,\qquad  \mathcal{W}_\mathcal{C} \,\approx\, e^{\,\mp \frac{2\pi i}{N}}  \;,
\end{equation}
which follows from the property (see Appendix \ref{App:weights})
\begin{equation}
  \label{cvw}
  e^{ i\, 2\pi \beta_i \cdot T} \,=\; e^{-i\:\!\frac{2\pi}{N}}\, \mathbb{1}\;.
\end{equation}
That is, the contribution to $\mathcal{W}_{\mathcal{C}}$ is that of an elementary center vortex (center charge $k=+1$ or $-1$), but here it is due to a collimated Cartan flux. Furthermore, note that the $N$ different Cartan fluxes $2\pi \beta_i \cdot T$, $i=1, \dots, N$ correspond to one and the same center element.

\subsection{Cartan-Based DeGrand-Toussaint Method}
\label{Cartanbased}

We now discuss monopoles in $SU(N)$, $N \geq 3$, as motivation for our proposed
detection method. The initial difficulty resides in the fact that the Cartan 
subalgebra has more than one dimension.
Let us consider, as before, the phases 
associated with diagonal elements of the projected link.
As noted above, such phases are not so simply connected 
with the variables $\theta_\mu(x)$ for $N\geq 3$. We therefore call them 
$\phi_\mu^{(i)}(x)$, $i=1,\cdots,N$.
Usually \cite{Stack:2002ysv},
after fixing to the MAG, the diagonal part of the
gauge links is extracted as in the $SU(2)$ case, by the maximization of the
overlap of the diagonal field $C_\mu(x)$ with the complete link variable
$U_\mu(x)$.
More precisely, in $SU(3)$, if we write
\begin{equation}
\vspace*{2mm}
\label{abelianprojectedSU3}
    C_\mu(x)\,=\, \left(\begin{array}{ccc}
         e^{i\phi^{(1)}_\mu(x)}& 0  & 0 \\
         0 & e^{i\phi^{(2)}_\mu(x)}  & 0  \\
         0 & 0 &  e^{i\phi^{(3)}_\mu(x)}   \end{array}\right) \,,
\qquad \phi^{(1)}_\mu,\;\phi^{(2)}_\mu,\;\phi^{(3)}_\mu\,\in\,[0,2\pi)\;,
\end{equation}
one maximizes ${\rm Re}\text{Tr}\,[U_\mu(x)\,C_\mu^\dagger(x)]$
with respect to the angle variables $\phi_\mu^{(i)}$, with the constraint
$\phi^{(1)}_\mu + \phi^{(2)}_\mu + 
\phi^{(3)}_\mu  =0\;\text{ mod }2\pi\,$.\footnote{
\label{symmetric_method}
Note that, in Ref.~\cite{Stack:2002ysv}, the maximization
procedure was found to be equivalent to simply extracting the phases of the
three diagonal elements of $U_\mu(x)$ and imposing the constraint on their
sum, i.e.\ considering each diagonal phase minus the average of the three
diagonal phases. This is the procedure we implement in our comparative 
study in Section \ref{results}.
}
(In \cite{Sakumichi:2014xpa,Suganuma:2018rvq},
a similar maximization was done with the
two Cartan components of $C_\mu$ as parameters.)

We then need a new prescription to compute the monopole currents.
For example, in Refs.~\cite{Stack:2002ysv,Suganuma:2018rvq}, the
DeGrand-Toussaint method described in the previous section was
applied independently to each one of the three fluxes 
$f^{(i)}_{\mu\nu}$ associated with $\phi_\mu^{(i)}$.
The measured monopoles would then have a charge with three integer 
components $(j^{(1)}_\mu,j^{(2)}_\mu,j^{(3)}_\mu)$. 
If we subdivide the space of the three phases into cubic cells of side
$2\pi$, then, extending the prescription in Eq.\,\eqref{HDecomposition},
one considers an ``initial'' point
$(f^{(1)}_{\mu\nu},f^{(2)}_{\mu\nu},f^{(3)}_{\mu\nu})$ in this space 
and takes it to a point
$(\bar{f}^{(1)}_{\mu\nu},\bar{f}^{(2)}_{\mu\nu},\bar{f}^{(3)}_{\mu\nu})$,
inside the unitary cubic cell centered at the origin. The path is, 
of course, a set of discrete steps of size $2\pi$ along each of the three 
directions. Note that the three flux components are treated as independent,
but are in fact correlated.
We refer to this as the 
{\em diagonal}
procedure.

\begin{figure}[h]
    \centering
    \includegraphics[width=0.5\linewidth]{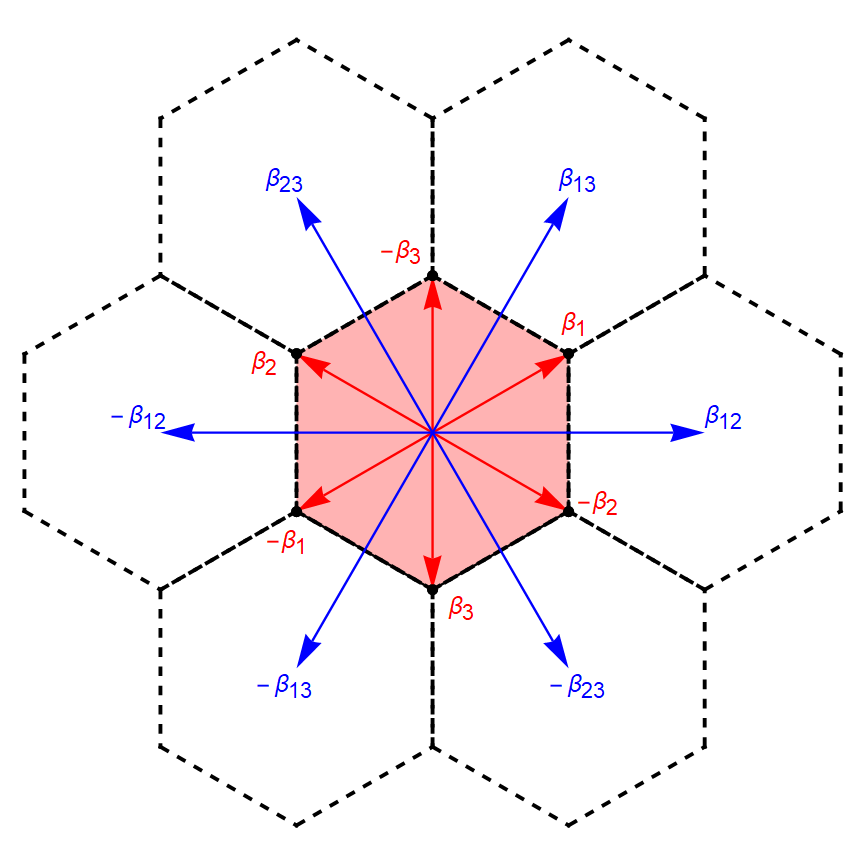}
    \caption{A graphical representation of the roots and defining weights of $SU(3)$.}
    \label{latt3}
\end{figure}

An alternative and non-equivalent way to proceed is to work with the natural
parametrization of Cartan fluxes, given in 
Eqs.\,\eqref{algebraicparametrization} and \eqref{algebraicplaquette}. 
This will lead to a mathematically sound
extension  of the procedure carried
out for the $U(1)$ and $SU(2)$ groups, emphasizing the main structures of the
Cartan subalgebra of $\mathfrak{su}(N)$. We refer to this approach as the
{\em Cartan-based procedure}.
In order to find the correct representation for a Dirac string in this 
language, let us consider a weight of the adjoint representation
$\alpha_{ij}= \omega_i - \omega_j$, cf.\ Eq.\,\eqref{difw}, and take
\begin{equation}
f_{\mu \nu} \,=\, 2\pi\,\beta_{ij}\,,\qquad 
\beta_{ij} \,=\, 2N\, \alpha_{ij} \;.
\label{efeal}
\end{equation}
Then we get, from the mathematical identity in Eq.\,\eqref{cvw}, that
\begin{equation}
    e^{ i\!\: f_{\mu \nu} \!\:\cdot\, T} \,=\;
    e^{i\, 2\pi\, \beta_{ij} \!\:\cdot\, T}=\, 
    e^{i\, 2\pi\, 2N\, \omega_i\!\: \cdot\, T}\,
    e^{-i\, 2\pi\, 2N \,\omega_j \!\:\cdot\, T} \,=\, \mathbb{1} \;.
    \label{dw}
\end{equation}
Note that this is true for any $i,j$.
The tuples $\beta_i$ and $\beta_{ij}$ are
called ``magnetic'' weights of the defining and adjoint representation, 
respectively \cite{Nyuts1977}. 
A depiction of them for $SU(3)$ is shown in Fig.\,\ref{latt3}.

\begin{figure}[h]
    \centering
    \includegraphics[width=0.5\linewidth]{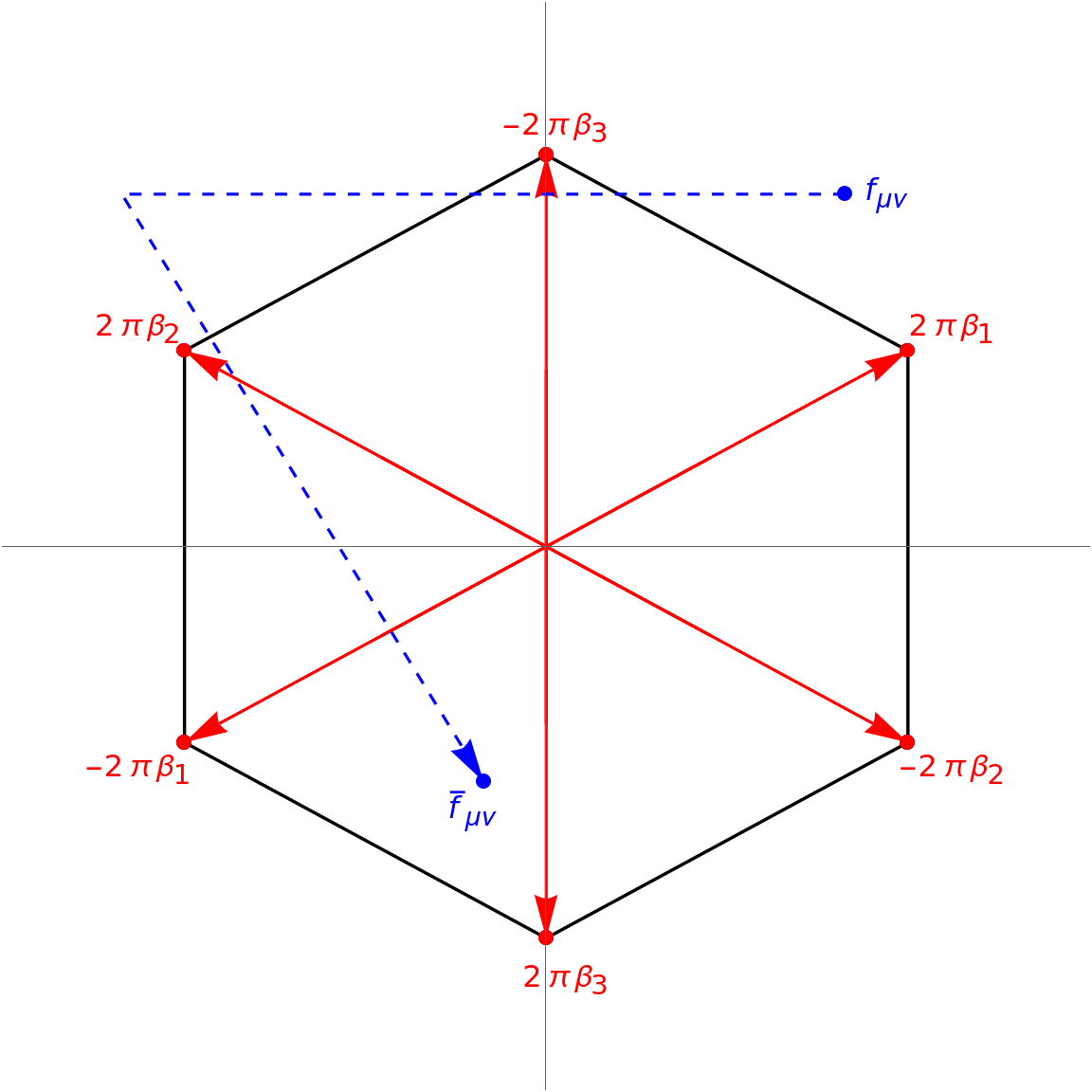}
    \caption{A point outside the fundamental hexagon is taken inside by a
     number of discrete steps along the direction of roots.}
    \label{PathtoHexagon}
\end{figure}

Thus, shifting a Cartan phase by elements of the lattice
generated by integer
combinations of $2\pi\,\beta_{ij}\cdot T$ plays a role analogous to the 
$2\pi$ phase periodicity of link variables in the compact Abelian case, 
Eq.~\eqref{HDecomposition}.
In particular, this clarifies how the DeGrand-Toussaint method can be 
extended\footnote{It is worth noting that this manner of implementing the 
DeGrant-Toussaint algorithm was first introduced in a previous work 
\cite{Bonati:2013bga}, but not in a way that assigns roots to the
detected monopoles, as we will be doing soon. }
using the parametrization in Eq.\,\eqref{algebraicparametrization}: the
appropriate terms that can be subtracted from the exponent in 
Eq.\,\eqref{algebraicplaquette} in order to find $\bar{f}_{\mu\nu}$ are
$2\pi\beta_{ij}$. It is worth noting that not all of them are
independent and, thus, only simple roots, with $j=i+1$, need 
to be considered.
Then, the natural way to define the DeGrand-Toussaint prescription in $\mathfrak{su}(N)$ is
\begin{equation}
\label{HDecomposition2}
    f_{\mu\nu} \,=\, \bar{f}_{\mu\nu} + 2\pi\,  l_{\mu \nu}\;,
\end{equation}
where $l_{\mu \nu}$ belongs to the lattice generated by $\beta_{ij}$ and 
is chosen such that the tuple $\bar{f}_{\mu\nu}$ lies in a unit cell,
that is, a region whose translations by the lattice vectors cover 
$\mathbb{R}^{N-1}$ exactly once (up to boundaries). For $\mathfrak{su}(3)$,
from Fig.~\ref{latt3}, we clearly observe two main properties. First, this
region can be chosen as the central hexagon (in red), scaled by a factor of
$2\pi$. Second, the set of all points $b \in \mathbb{R}^2$ in this hexagon is
characterized by having projections along the directions 
$\beta_{ij}/|\beta_{ij}|$ with modulus at most $|\beta_{ij}|/2$:  
\begin{equation}
b \cdot \frac{\beta_{ij}}{|\beta_{ij}|} 
\,\in\, \biggl[-\frac{|\beta_{ij}|}{2}, \,+\frac{|\beta_{ij}|}{2}\biggr] \;.
\end{equation}
In fact, this condition also determines the unit cell for general $N$, where
$b \in \mathbb{R}^{N-1}$.
From Eq.~\eqref{efeal}, and using our conventions, we have 
$|\alpha_{ij}| = 1/\sqrt{N}$, see Eq.\,\eqref{roots3} for $\mathfrak{su}(3)$.
Thus, the unit cell is given by
\begin{equation}
\frac{1}{2}\;\bar{f}_{\mu \nu} \cdot \alpha
\;\in\, [-\pi , +\pi] \;,
\label{condi}
\end{equation}
for every positive root $\alpha$. 
Therefore, if a plaquette variable $f_{\mu\nu}$ is a point outside the 
fundamental hexagon, it is taken inside the unit cell, 
i.e.\ it is moved until the condition in Eq.\,\eqref{condi} is met, by a 
set of discrete steps along the directions $\beta_{ij}$. This is shown 
in Fig.\,\ref{PathtoHexagon}.

Just as in the Abelian or $SU(2)$ case, the tensor $2\pi\,l_{\mu \nu}$ is the
lattice representation of a Dirac surface. Due to Eqs.\,\eqref{efeal} and 
\eqref{dw}, it has no effect on (projected) Wilson loops or on the 
Wilson action. In addition, when the sum of the fluxes $2\pi l_{\mu \nu}$ through all
the faces of a cube is nonzero, a monopole is detected. Of course, the
sum of the fluxes $\bar{f}_{\mu \nu}$ will then be nonzero and opposite. The plaquettes
with $l_{\mu \nu} \neq 0$ can be changed by a gauge transformation, while the
monopole locations cannot (see Section \ref{compa}). Thus, monopoles are the
only physical degrees of freedom associated with $l_{\mu \nu}$.  

Interestingly, the unit cell defined by Eq.\,\eqref{condi} is a regular 
higher-dimensional polytope with vertices at $\pm 2\pi \beta_i$. Thus, although this
construction is oriented toward detecting monopoles, it is closely related to
the fluxes of elementary center vortices, as discussed in Section \ref{f-col}.
This, in turn, is related to the possibility of nonoriented collimated center
vortices (see Section \ref{non-corr}). Of course, for $N=2$, the  vertices
correspond to Cartan fluxes at $\pm 2\pi\, 2\alpha_1 T_1 =\pm \pi \sigma_3$,
cf.\ Eqs.\,\eqref{T1} and \eqref{alphasu2}. These are the center-vortex fluxes
around monopoles in $\mathfrak{su}(2)$, whose collimation has already been 
studied (see Section \ref{current-d}).

\subsection{Comparison of the Detection Methods}
\label{compa}

At this point, it is enlightening to compare the above described
{\em diagonal} 
and \emph{Cartan-based} procedures. 
Regarding the 
diagonal procedure, notice that the three phases in 
Eq.\,\eqref{algebraicparametrization} are not independent, but
this does not imply a strong dependence among the monopole
currents $j_\mu^{(i)}$, which do not need to sum to $0$ themselves.
This can lead to unphysical currents, even though it has been argued that they have no effect on average \cite{Stack:2002ysv}.
In any case, one should determine
the best way to extract physical information from the fluxes in the 
exponent of the Abelian-projected variables for $N\geq 3$.

In order to discuss this in more detail, let us introduce a common notation.
Both methods start from the Cartan-projected link variables parametrized as
in Eq.\,\eqref{algebraicparametrization} or Eq.\,\eqref{abelianprojectedSU3},
that is, with an exponent in the Cartan subalgebra, which we write as
\begin{equation}
    C_\mu(x) \,=\, e^{i \Phi_\mu(x)}\;, \qquad {\rm Tr}(\Phi_\mu) = 0 \;.
\end{equation}
The corresponding projected plaquette variables are
\begin{equation}
C_{\mu\nu}(x) \,=\, e^{i F_{\mu\nu}(x)}\;, \qquad 
F_{\mu \nu} = \partial_\mu \Phi_\nu - \partial_\nu \Phi_\nu\;.
\end{equation}
In principle, the physical content is encoded in the group-valued $C_{\mu\nu}$.
On the other hand, monopole detection is performed at the level of the phases
$F_{\mu\nu}$, which are in the Cartan subalgebra.
The generalized DeGrand-Toussaint procedure defined in both cases involves
a decomposition of the form
\begin{equation}
F_{\mu\nu} \,=\, \widebar{F}_{\mu\nu} + 2\pi\, N_{\mu\nu} \;,
\label{decompositioncube}    
\end{equation}
where $\widebar{F}_{\mu \nu}$ lies in a given domain and $N_{\mu \nu}$ is a 
diagonal matrix with integer entries, which may be constrained or 
unconstrained. Indeed, in the Cartan-based case, the traceless constraint is
resolved by $F_{\mu \nu} =  f_{\mu \nu}\cdot T$ and the decomposition 
$F_{\mu \nu}= \bar{f}_{\mu \nu} \cdot T + 2\pi\, l_{\mu \nu} \cdot T$, as the
second term is equal to ($2\pi$ times) a general traceless 
$N_{\mu \nu}$.\footnote{Note that $l_{\mu \nu}$ belongs to the lattice
generated by $\beta_{ij}$ and, from Eq.\,\eqref{dw}, $\beta_{ij}\cdot T$
necessarily has integer entries.} 

In both methods, the monopole currents are identified through 
\begin{equation}
j_\mu \,=\, - \partial_\nu \widetilde{N}_{\mu \nu} \,, \qquad  
\widetilde{N}_{\mu \nu} \,=\, \frac{1}{2}\, \epsilon_{\mu\nu\rho\sigma}
\, N_{\rho \sigma} \;,
\end{equation}
which is the analogue of the Abelian case in Eq.\,\eqref{mmu}.
Now, in the
diagonal
method, it is not always possible to find a traceless 
$N_{\mu \nu}$ such that $\widebar{F}_{\mu \nu}$ lies within the 
``cube'' defined by constraining all three phases to
the interval $[-\pi,\pi]$. 
Since $F_{\mu \nu}$ is Cartan-valued, the possibilities split into two 
cases: either both $\widebar{F}_{\mu \nu}$ and $N_{\mu \nu}$ lie in 
the Cartan subalgebra, or both lie outside it. 
In the first case, it is interesting to note that $\widebar{F}_{\mu \nu}$
coincides for both the 
diagonal
and the Cartan-based procedures. 
If we show that a traceless $\widebar{F}_{\mu\nu}$, with parameters in the cube,
satisfies the conditions in Eq.\,\eqref{condi} to be inside the fundamental
hexagon, then it must be true that 
$\widebar{F}_{\mu\nu}=\bar{f}_{\mu\nu}\cdot T$,
since both decompositions are unique. 
In terms of $\widebar{F}_{\mu\nu}$, these conditions are
\begin{equation}
N\,\text{Tr} \left(\widebar{F}_{\mu\nu}\, \alpha\cdot T\right)
\,\in\,[-\pi,\pi]  \;,
\end{equation}
for every root $\alpha$, which are differences of weights: 
$\alpha=\omega_i-\omega_j$, cf.\ Eq.\,\eqref{difw}. Then, by using 
Eqs.\,\eqref{wfun} and \eqref{fundwei}, it can be shown that
\begin{equation}
N\,\text{Tr} \left(\widebar{F}_{\mu\nu}\, \alpha\cdot T\right) 
\,=\,\frac{\widebar{F}_{\mu\nu}\vert_{ii}-\widebar{F}_{\mu\nu}\vert_{jj}}{2}\;,
\end{equation}
where $\widebar{F}_{\mu\nu}\vert_{ij}$ denotes a matrix
element of $\widebar{F}_{\mu\nu}$.
Consequently, whenever the diagonal components of $\widebar{F}_{\mu\nu}$ are in 
the interval $[-\pi,\pi]$ and sum to zero, the corresponding 
$\bar{f}_{\mu \nu}$ is in the fundamental hexagon.
Of course, if this applies to all the plaquettes of a cube, then the monopole
detection will coincide on this cube.  On the other hand, because of the 
second possibility, when following the diagonal method, 
monopoles may carry unphysical fluxes outside the Cartan algebra.

Our proposal directly targets the physical sector. As a side remark, note that
the phases of the link variables in $\Phi_\mu$ are not unique. One may shift
\begin{equation}
\Phi_\mu(x) \;\to\; \Phi'_\mu (x) = \Phi_\mu(x) + 2\pi\, L_\mu(x) \;,
\end{equation}
where $L_\mu(x)$ is a diagonal matrix with integer entries. In both the constrained and unconstrained cases, this integer phase 
transformation leaves $C_\mu$ and $C_{\mu \nu}$ unchanged, while inducing
\begin{equation}
F_{\mu\nu} \;\to\; F'_{\mu \nu} = F_{\mu\nu} + 2\pi\, (\partial_\mu L_\nu - \partial_\nu L_\mu) \;.
\label{L-t}
\end{equation}
As a consequence, $\widetilde{N}_{\mu \nu}$ is changed by the addition of 
$\epsilon_{\mu \nu \rho \sigma} \partial_\rho L_\sigma$, which is localized
on closed surfaces. In the constrained case, this corresponds to large gauge
transformations that change the Dirac worldsurfaces, while leaving 
$\widebar{F}_{\mu \nu}$ invariant. In the unconstrained case, to move the 
open surfaces encoded in $\widetilde{N}_{\mu \nu}$, the Cartan gauge subgroup
should be enlarged to allow for large transformations carrying three 
independent $U(1)$ fluxes.

Finally, since our method focuses on the defining properties of the root lattice, cf.\ Eq.\,\eqref{condi}, which characterizes the $\mathfrak{su}(N)$ 
Lie algebra, it naturally encodes non-Abelian information. Moreover, it 
enables the detection of additional correlated objects in the Yang-Mills
vacuum, as discussed in the next section.

\subsection{Nonoriented Center Vortices and Other Correlations in $SU(N)$} 
\label{non-corr}

\begin{figure}[h]
    \centering
    \includegraphics[width=0.6\linewidth]{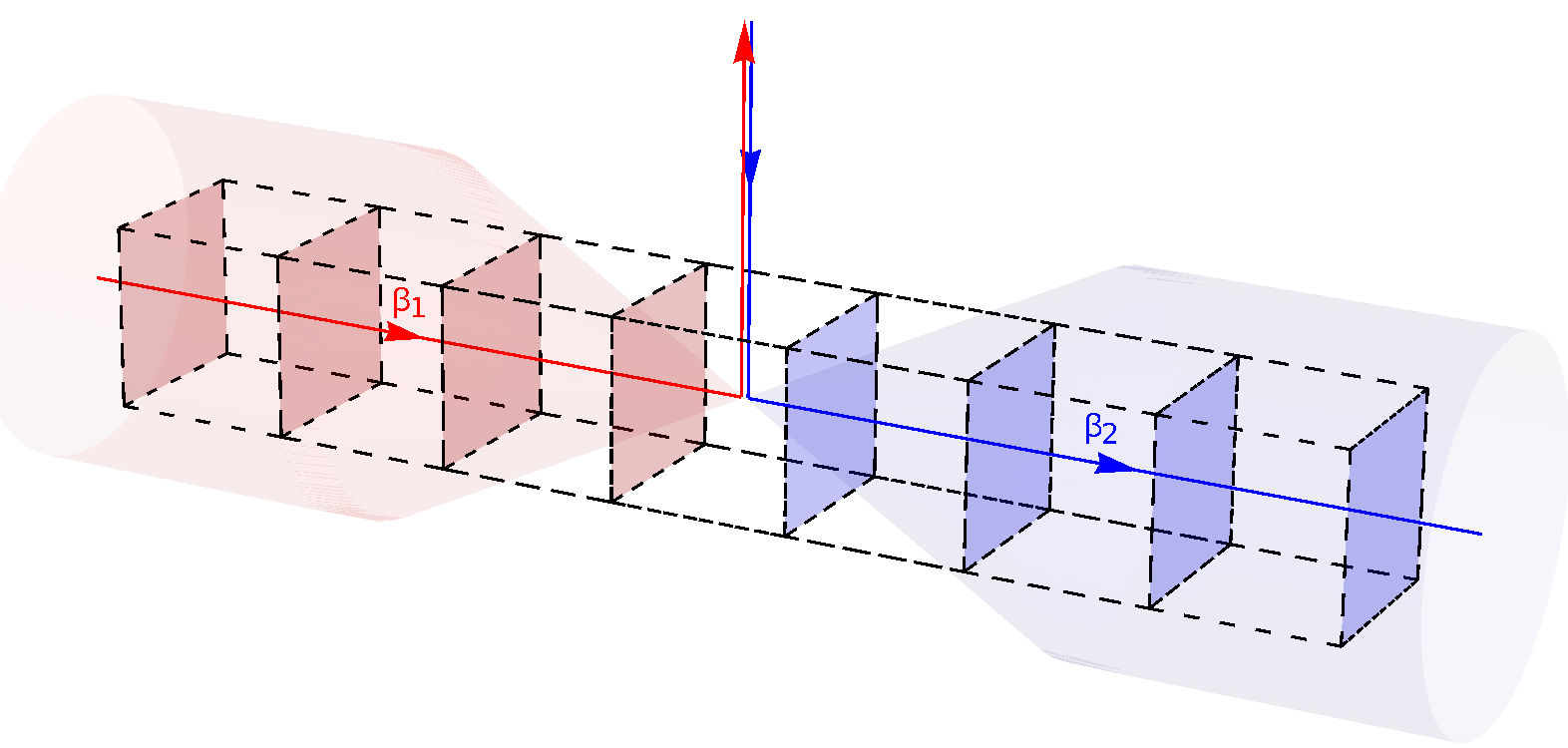}
    \caption{A nonoriented center vortex changing the flux orientation in the Lie algebra, from $\beta_1\cdot T$ to $\beta_2\cdot T$. The monopole can be detected at an endpoint  of a Dirac string carrying charge $\beta_{21}=\left(\beta_2-\beta_1\right)$. The squares do not necessarily represent elementary plaquettes; they have a size comparable to that of the collimated vortices. }
    \label{nono}
\end{figure}

Let us consider the elementary fluxes $2\pi \beta_{ij} \cdot T$ of monopoles
on the Abelian-projected lattice, where $\beta_{ij} = \beta_i - \beta_j$ and
we refer to Eq.\,\eqref{betadef}. It is natural to expect the emanating flux
to be collimated along a pair of attached center-vortex lines. Indeed, as 
discussed in Section \ref{f-col}, total fluxes $2\pi \beta_i \cdot T$ 
correspond to those carried by elementary center vortices. 
Such center-vortex/monopole chains are also known as nonoriented vortices
and are represented in Fig.\,\ref{nono}. On one side of the monopole, the
Cartan flux, with a distribution that is locally close to
$a_{\mu \nu}\,\beta_1 \cdot T$, changes across the monopole to a distribution
given locally by $a_{\mu \nu}\,\beta_2 \cdot T$. Note that this change of
orientation occurs in the Lie algebra. This is clearly the case for $N \geq 3$.
In $SU(2)$, where the elementary fluxes are $\pm \pi\,\sigma_3$, this change
may be confused with a change in spatial orientation, in which $a_{\mu \nu}$
changes sign from one side to the other.

\begin{figure}[h]
    \centering
    \includegraphics[width=0.55\linewidth]{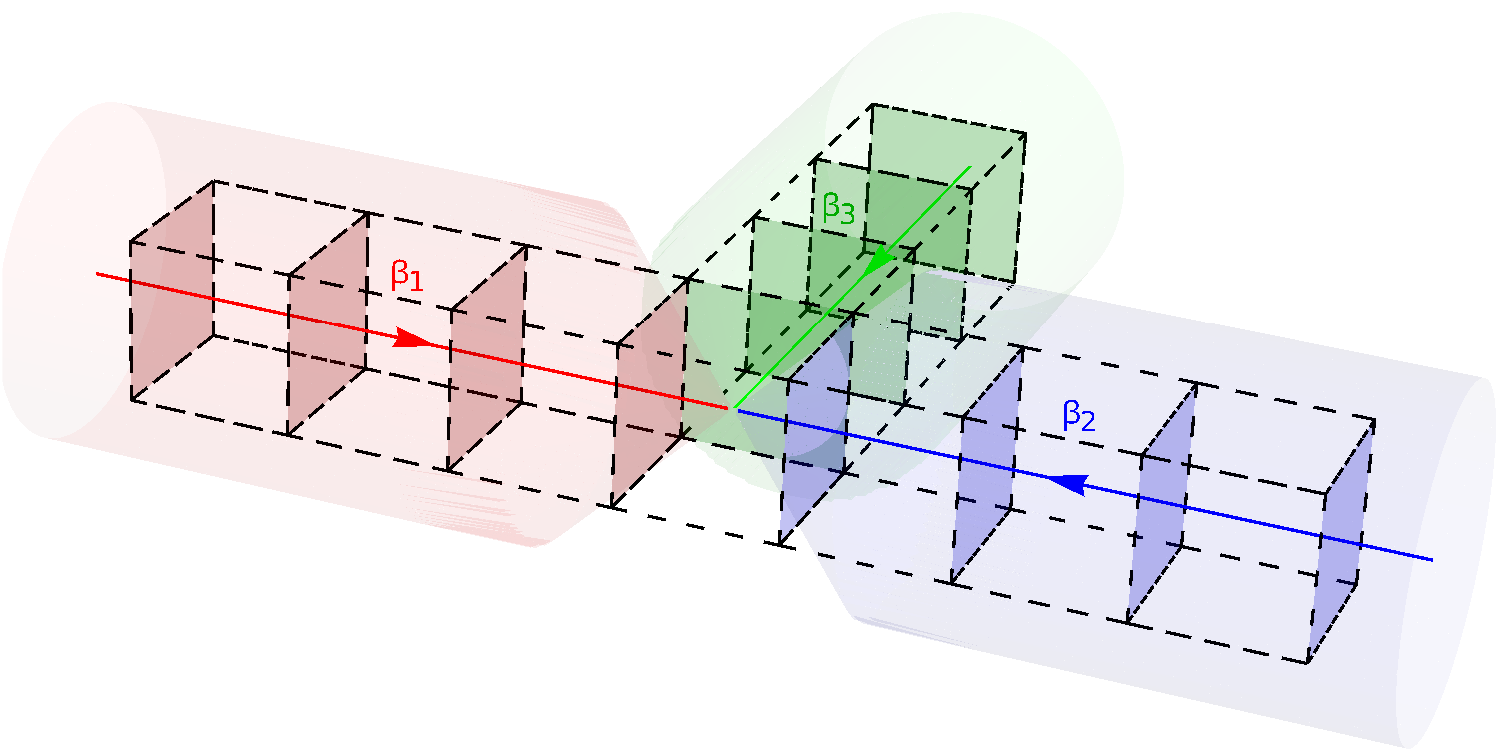}
    \caption{Three-vortex array with different charges. The matching point is not detectable by DeGrand-Toussaint methods, but can be identified as a $Z(3)$ monopole via center projection using the IMC. The squares represent regions with size comparable to that of the collimated vortices.}
    \label{Nmatching}
\end{figure}

The characterization of Cartan fluxes may also be helpful to study flux
distributions around branching points in networks of vortex guiding centers,
as detected by the procedure based on fixing the IMC gauge. These points have
been detected in the DMC gauge by identifying cubes containing a $k=3$ center
monopole, in $SU(3)$, and they constitute fundamental building blocks of the
network \cite{Biddle:2019gke}. With our method, one could also address the
question of the relative frequencies of different Cartan distributions that
produce the same $Z(3)$ center monopole at a junction. 
A possible configuration is depicted in Fig.\,\ref{Nmatching}, where three
fluxes characterized by three different magnetic weights
$\beta_1, \beta_2, \beta_3$ are matched at a point. This is possible because
of the relation $\beta_1 + \beta_2 + \beta_3 = 0$. 
Such occurrence should be studied by first using the IMC gauge since, at the
level of the Abelian-projected lattice, there would be no Cartan monopole at
the matching point, that is, no Dirac string in the lattice data. 
On the other hand, a $Z(3)$ center monopole could also coincide with a Cartan
monopole, which could be detected by any of the generalized DeGrand-Toussaint
methods described above. For example, this would be the case when three
collimated center vortices carrying flux $2\pi \beta_1 \cdot T$ enter a 
monopole, which can be detected by a Dirac string carrying flux, with magnetic
charge $3\beta_1 = \beta_{12} + \beta_{13}$.

At present, there are no lattice studies of the above-mentioned flux
distributions. Some indirect information about these correlations can be 
inferred from the results in Ref.~\cite{Stack:2002ysv} 
(see Section~\ref{current-d}). We first note the relatively large fraction
($16\%$) of monopoles not sitting on center vortices, as compared to $3\%$
in $SU(2)$. We speculate that this may be related to spurious monopoles
carrying $U(1)^3$-like fluxes that are not genuine Cartan fluxes.
The $74\%$ fraction of Abelian-projected monopoles correlated with a pair of
$P$-plaquettes would in principle be associated with nonoriented
configurations. In addition, about $7\%$ of Cartan monopoles would coincide
with $Z(3)$ monopoles. However, no information is available to compare the
fractions of $Z(3)$ monopoles that do and do not coincide with a Cartan
monopole. There is also the possibility of configurations where monopole
worldlines carrying different adjoint charges meet at points on the 4D 
lattice. Such configurations are allowed by flux conservation. For example,
three worldlines carrying adjoint magnetic fluxes $\beta_{12}$, $\beta_{23}$,
and $\beta_{31}$ can meet at a point, since
$\beta_{12} + \beta_{23} + \beta_{31} = 0$.

In Refs.~\cite{Oxman:2018dzp,Oxman:2018nqe,Junior:2019fty,Junior:2022bol,Junior:2024urr}, all
these elements ---namely, oriented and nonoriented center vortices, the 
possibility of center-vortex $N$-matching, as well as monopole fusion--- were extensively explored (for a comprehensive review, see 
Ref.~\cite{Junior:2021vpd}). 

In particular, Ref.~\cite{Oxman:2018dzp} proposed a mixed ensemble of oriented and nonoriented collimated fluxes as a mechanism for confinement. 
There, percolating oriented center-vortex worldsurfaces in 4D were effectively described by emergent gauge fields $\Lambda_\mu$, which represent the Goldstone modes in the center-vortex condensate. The nonoriented component gives rise to monopole scalar fields minimally coupled to $\Lambda_\mu$. In addition, the monopole condensate and monopole fusion lead to a phase with spontaneously broken symmetry. In the continuum, these are key ingredients for the generation of topologically stable confining flux tubes between quarks and for the confinement of valence gluons.

Along this line of research, Ref.~\cite{Junior:2022bol} proposed a vacuum wave functional \(\Psi[A]\) peaked at Abelian-projected center-vortex configurations. Since this approach deals with gauge fields \(A(x)\) defined at a given time slice, the configurations become networks of center-vortex lines in \(\mathbb{R}^3\). This allowed the use of a mechanism similar to that implemented in 3D spacetime in Ref.~\cite{Junior:2019fty}, but in the 3+1-dimensional theory. The resulting scaling properties of the confining string tension coincide with those derived in Ref.~\cite{Oxman:2018wpb}, based on the 4D effective model of Ref.~\cite{Oxman:2018dzp}. The Abelian-projected configurations were also analyzed in  Ref.~\cite{Junior:2024urr}, but at the level of the 4D partition function, where the sum over center-vortex worldsurfaces was formulated using a lattice matrix model.

\section{Maximally Abelian Gauge}
\label{MAG}

We now describe the gauge-fixing step, namely our prescription for
fixing the MAG. We begin by describing the gauge condition in general,
detailing our lattice implementation in Section \ref{MAGlatt}.
We motivate the use of the Weyl-symmetric form of the gauge condition
in Section \ref{WeylSymm}.

In the Maximally Abelian Gauge (MAG) \cite{tHooft:1981bkw}, the
off-diagonal components of the gauge field are minimized, effectively 
forcing these degrees of freedom to become massive 
and decouple at low energies, see e.g.\ \cite{Dudal:2005bk}.
In turn, the Abelian-like part is maximized, highlighting the Abelian degrees
of freedom in the theory, believed to drive confinement.
This is done through the minimization of the functional
\begin{equation}
	\label{continuum MAG}
	F=\int\,d^4x\;\sum_\mu\sum_{a=N}^{N^2-1}\,{A_\mu^a(x)}^2
\end{equation} 
of the gauge fields $A_\mu^a(x)$, where $\mu$ is the Lorentz index 
and $a=1,\dots,N^2-1$ is the color index associated with the adjoint 
representation of $\mathfrak{su}(N)$. 
We adopt the notation (see Appendix \ref{Cartan-basis}) in which the 
first $N-1$ color indices
$q=1,\dots,N-1$ refer to the diagonal part of the algebra, while
$a=N,\dots,N^2-1$ refer to the off-diagonal part. For instance, the index
$q=1$ is associated with the Pauli matrix $\sigma_3$ for $SU(2)$, while 
$q=1,2$ are respectively associated with the Gell-Mann matrices $\lambda_3$
and $\lambda_8$ for $SU(3)$.
Minimizing this functional yields the MAG gauge-fixing conditions. 
For the $SU(2)$ case, these are 
\cite{Stack:1994wm}
\begin{equation}
\label{thooft abelian gauge conditions}
\partial_\mu\,A_\mu^\pm(x)\,\pm i\,A_\mu^1(x)\,A_\mu^\pm(x)=0\;,
\end{equation} 
where summation of repeated indices is implied and
\begin{equation}
A_\mu^\pm(x) \equiv A_\mu^2(x)\pm\,i\,A_\mu^3(x)\;.
\end{equation} 
In the $SU(3)$ case, the same conditions may be imposed for each $SU(2)$ 
subgroup. 
In general, the condition may be written \cite{Dudal:2005bk} as
\begin{equation}
D_\mu^{a b} A^{\mu\,b}\;=\; 0\,,\quad a,b\neq 1,...,N-1\;,
\end{equation}
where $D_\mu$ is the covariant derivative acting on the gauge field.
We therefore have a nonlinear gauge condition and a residual 
$U(1)^{N-1}$ symmetry.

\subsection{Lattice Implementation}
\label{MAGlatt}

On a Euclidean lattice, the MAG conditions are reproduced by finding the gauge
configuration corresponding to the maximum of the functional\footnote{In 
Ref.~\cite{Bonati:2013bga}, the authors argue in favor of a different 
implementation of the MAG, although the Cartan-based procedure is 
qualitatively independent of this choice. 
Nevertheless, as explained in Section \ref{WeylSymm}, the usual definition
of the MAG is especially adequate for our work.}
\begin{equation}
	\label{lattice mag}
  {\cal E}=\sum\limits_{x,\mu,q}\text{Tr }
  \big[U_\mu(x)\,T_q\,U^\dagger_\mu(x)\,T_q\big]\;
\end{equation}
where $U_\mu(x)\in SU(N)$ is the lattice gauge link variable and $T_q$ are 
the Cartan generators, i.e. the diagonal part of the algebra (see
Appendix \ref{Cartan-basis}). As above, we 
have $q=1,\dots,N-1$. This is the gauge-fixing condition adopted in our work.
Notice that, in the limit of small lattice spacing, the above functional
reduces to (minus) that of Eq.\,\eqref{continuum MAG}.

Therefore, our problem is, given a thermalized gauge configuration 
$\{U_\mu\}$, to find the local transformation $\{g\}$,
$g(x)\in SU(N)$, that will bring the link variables to 
$\{U^g_\mu\}$, the configuration corresponding to the extremum of 
the functional in Eq.\,\eqref{lattice mag}. 
This is achieved with an iterative process, in which we find the 
gauge transformation $g(x)$ at a given lattice site $x$ that maximizes
${\cal E}(x)$ ---defined by the expression in Eq.\,\eqref{lattice mag} 
considering only the terms involving $g(x)$--- while, at all other sites, 
the associated links are kept unchanged.
We then update the links connected to $g(x)$, i.e.\ $U_\mu(x)$ and 
$U_\mu(x-\hat{\mu})$, and move on to the next site, applying the procedure
to all lattice sites. Since the functional in Eq.\,\eqref{lattice mag} is 
bounded from above, by repeating this process many times, we expect to 
reach a (local) maximum of the gauge functional.
More precisely, we have
\begin{eqnarray}
\label{MAGfunctional}
       {\cal E}(x) & = &\sum_{\mu,q} \text{Tr}\,\Big[
  g(x)\,U_\mu(x)\, T_q \,U^\dagger_\mu(x)\,g^\dagger(x)\,T_q \,+\,
  g(x)\,U_\mu^{\dagger}(x -\hat{\mu})\, T_q\,U_\mu(x-\hat{\mu})\,
  g^\dagger(x)\,T_q \Big] \nonumber \\[2mm]
       & = &\sum_q \text{Tr}\,\big[ n_q(x)\,A_q(x) \big]\;,
       \label{mean}
    \end{eqnarray}
where we have defined\footnote{Note that $n_q(x)$ and $A_q(x)$ are elements
of the Lie algebra $\mathfrak{su}(N)$.}
    \begin{align}
n_q(x) = g^{\dagger}(x)\,T_q \, g(x) \,,\quad
A_q(x) =  \sum_\mu \bigg[U_\mu(x)\, T_q \,U^\dagger_\mu(x) 
\,+\, U^\dagger_\mu(x-\hat{\mu})\, T_q\,U_\mu(x-\hat{\mu})\bigg] \;.
\label{nqAq}
    \end{align}
Then, the problem of maximizing ${\cal E}(x)$ is
that of finding $g(x)$ such that $n_q(x)$ is oriented as much as possible
along the ``direction'' of $A_q(x)$, on average (when summing over
$q$), cf.\,Eq.\,\eqref{mean}.
Equivalently, we may maximize the functional
${\cal E}(x)$ directly, as described below.

\vskip 3mm 
For $SU(2)$, the contribution to the MAG 
functional in Eq.\,\eqref{MAGfunctional} at a site $x$ takes the form
\begin{equation}
\label{MAGfunctionalSU2}
{\cal E}(x)\;=\frac{1}{8}\;
\sum_\mu\,\text{Tr}\,\bigg[g(x)\,U_\mu(x)\,\sigma_3\,U^\dagger_\mu(x)
\,g^\dagger(x)\,\sigma_3 \,+\, g(x)\,U^\dagger_\mu(x-\hat{\mu})\,
\sigma_3\,U_\mu(x-\hat{\mu})\,g^\dagger(x)\,\sigma_3\bigg]\;,
\end{equation}
where $\sigma_i$, $i=1,2,3$ are the Pauli matrices.
It is natural to treat the functional as the interaction of an effective 
spin $\,s(x)=g^\dagger(x)\,\sigma_3\,g(x)$ with an external magnetic field 
$\,h(x)=\sum_\mu \big[U_\mu(x)\,\sigma_3\,U^\dagger_\mu(x)+
U^\dagger_\mu(x-\hat{\mu})\,\sigma_3\,U_\mu(x-\hat{\mu})\big]$,
respectively associated with $n_q(x)$ and $A_q(x)$ in Eq.\,\eqref{nqAq}.
In this way, the $SU(2)$ MAG functional assumes an identical form to that of 
the lattice Landau gauge functional, allowing treatment using spin-glass 
techniques
\cite{Mendes:2006kc,Cucchieri:2003fb,Cucchieri:1995pn}.
Alternatively, we may
work with the component form of the MAG functional
\cite{Stack:2002ysv}.

Let us write
\begin{equation}
g(x) \,=\, g^4(x)\,\mathbb{1}+ig^1(x)\,\sigma_1+ig^2(x)\,\sigma_2\;,
\label{su2param}
\end{equation}
where $g^3(x)=0$, since a residual $U(1)$ symmetry is left--- i.e.\ we 
cannot fix the phase of the diagonal elements--- and the links as
\begin{equation}
U_\mu(x) \,=\, u_\mu^4(x)\,\mathbb{1}+i\,u_\mu^a(x)\,\sigma_a
\,\equiv\, u_\mu^4(x)\,\mathbb{1}+i\,\vec{u}_\mu(x)\cdot\vec{\sigma} \;,
\end{equation}
where the color index $a$ is summed over, and we defined three-vectors
with components $1,2,3$ of the appropriate quantities.
Thus, the functional at $x$ assumes the form
\begin{eqnarray}
{\cal E}(x) &=& \frac{1}{8}\sum_\mu\sum_{i=1,2}
\bigg\{g^4 u_\mu^i(x)+u_\mu^4(x)\,g^i \,-\,
\Big[\vec{g}\times\vec{u}_\mu(x)\Big]_i \bigg\}^2 
\,+\, \frac{1}{8}\sum_\mu \Big[g^4 u_\mu^4(x)-\vec{g}\cdot\vec{u}_\mu(x)\Big]^2
\nonumber \\[2mm] & & 
\,+\,\bigg[U_\mu(x)\leftrightarrow U^\dagger_\mu(x-\hat{\mu})\bigg]
\,+\, \frac{\xi}{8}\,\sum_{i=1,2,4}\Big[(g^i)^2-1\Big]\;,
\end{eqnarray}
where we indicate the expression obtained from substituting 
$U_\mu(x)$ by $U^\dagger_\mu(x-\hat{\mu})$ 
in the first line of the above equation, and we have inserted a 
Lagrange multiplier to guarantee that $g(x)$ is 
still in $SU(2)$. The extremization conditions are then given by 
$\,\partial {\cal E}/\partial g^i=0$, yielding the eigenvalue equation
\cite{Schrock:2012fj}
\begin{equation}
\begin{pmatrix}
	-C & 0 & A \\ 0 & -C & B \\ A & B & C
\end{pmatrix}\,\begin{pmatrix}
g^1 \\ g^2 \\g^4
\end{pmatrix}=\;\xi\,\begin{pmatrix}
g^1 \\ g^2 \\g^4
\end{pmatrix}\;,
\label{eigenvalues}
\end{equation}
where
\begin{eqnarray}
A &=& 2\,\sum_\mu \Big[u_\mu^1(x)\,u_\mu^4(x)+u_\mu^2(x)\,u_\mu^3(x)\Big]
\,+\,\bigg[U_\mu(x)\leftrightarrow U^\dagger_\mu(x-\hat{\mu})\bigg]\;, \\[2mm]
B &=& 2\,\sum_\mu \Big[u_\mu^2(x)\,u_\mu^4(x)-u_\mu^1(x)\,u_\mu^3(x)\Big]
\,+\,\bigg[U_\mu(x)\leftrightarrow U^\dagger_\mu(x-\hat{\mu})\bigg]\;, \\[2mm]
C &=& \sum_\mu \bigg[\,\frac{1}{2}-
u_\mu^3(x)\,u_\mu^3(x)-u_\mu^4(x)\,u_\mu^4(x)\bigg]
\,+\,\bigg[U_\mu(x)\leftrightarrow U^\dagger_\mu(x-\hat{\mu})\bigg]\;.
\end{eqnarray}
The eigenvector corresponding to the largest eigenvalue is the one that maximizes the functional. Hence, the gauge-transformation matrix is given by
\begin{equation}
\label{su(2) mag solution}
g(x)\,=\,\mathcal{P}_{SU(2)}\left(A,\,B,\,0,\,C+\sqrt{A^2+B^2+C^2}\right)^T\;,
\end{equation}
where $\mathcal{P}_{SU(2)}$ means projection onto the $SU(2)$ subgroup. 
This solves the problem in the $SU(2)$ case.

For $SU(3)$, the MAG functional is more involved, due to the presence
of two diagonal matrices in the algebra, as mentioned above.
We have
\begin{equation}
{\cal E} \,=\, \frac{1}{12}\sum_{x,\,\mu}\,\text{Re Tr }
\Big[U_\mu(x)\,\lambda_3\,U^\dagger_\mu(x)\,\lambda_3
+U_\mu(x)\,\lambda_8\,U^\dagger_\mu(x)\,\lambda_8\Big]\;,
\label{SU3funct}
\end{equation}
where $\lambda_a$ ($a=1,\dots,8$) are the Gell-Mann matrices. 
In terms of components of the $U_\mu(x)$ matrix, the expression simplifies
and we have
\begin{equation}
{\cal E}\,=\, \frac{1}{6}\sum_{x,\,\mu}\,\left[\,\sum_{i=1}^3
\left|{U_\mu(x)}_{ii}\right|^2-1\,\right]\;,
\label{Ecomponents}
\end{equation}
where $U_\mu(x)$ must be thought of as $U^g_\mu(x)$, i.e.\ the 
transformed gauge variable.
Again, we apply an iterative procedure, sweeping over the sites and finding 
$g(x)\in SU(3)$ that gauge-fixes the links touched by it. At a given site $x$,
we search for the desired $SU(3)$ gauge transformation decomposing the matrix
into its $SU(2)$ subgroups. We denote the three $SU(2)$ subgroups by the tuple 
$(\alpha,\beta) \in \{(1,2),(1,3),(2,3)\}$, which allows us to write the matrix
elements of $g(x)\in SU(3)$ in terms of an effective $SU(2)$ matrix.
More precisely, for fixed $(\alpha,\beta)$, the matrix elements
${g(x)}_{ij}$ ($i,j=1,2,3$) may be parametrized by four real numbers
$r^k$ ($k=1,\dots,4$) as
\begin{eqnarray}
g^{\alpha\beta}_{ij} &=& (\delta_{i \alpha}\,\delta_{j \alpha}+
\delta_{i \beta}\,\delta_{j \beta})\,r^4 \,+\,
i\,(\delta_{i \alpha}\,\delta_{j \alpha}-
\delta_{i \beta}\,\delta_{j \beta})\,r^3
\,+\,(\delta_{i \alpha}\,\delta_{j \beta}-
\delta_{i \beta}\,\delta_{j \alpha})\,r^2
\nonumber \\[2mm]
& & 
\,+\,i\,(\delta_{i \alpha}\,\delta_{j \beta}+
\delta_{i \beta}\,\delta_{j \alpha})\,r^1
\,+\,\delta_{i \gamma}\,\delta_{j \gamma}\;,
\label{gsubgroup}
\end{eqnarray}
where we have defined $\gamma\equiv 6-\alpha-\beta$ and $\delta_{ij}$ is
the Kronecker delta. We thus have an embedded $SU(2)$ matrix described
by the form in Eq.\,\eqref{su2param}, subject to the
condition $\sum_k ({r^k})^2=1$. Also, we take $r^3=0$.
In this way, 
we have reduced our problem to that of finding an $SU(2)$ gauge transformation
that brings the subgroup to the MAG, for which we already have a solution.
Indeed, adopting the expression \eqref{gsubgroup} for the
gauge transformation in Eq.\,\eqref{Ecomponents} and maximizing the
functional restricted to the site $x$, we obtain again the set of conditions in
Eq.\,\eqref{eigenvalues}.
The coefficients $A,B,C$
are now given by the projection of the matrices onto the subgroup, 
and the gauge transformation for this subgroup is given by 
Eq.\,\eqref{su(2) mag solution}. 

Sweeping over the three $SU(2)$ subgroups
gives us the $SU(3)$ gauge transformation at this lattice site.
We apply the method just described to all lattice sites,
computing the $SU(3)$ MAG functional at the end of each sweep. When its 
variation is smaller than $10^{-9}$ for 50 consecutive times, we say that 
the procedure has converged and stop it. 
Actually, following the overrelaxation method, at each site $x$ we
apply ${g(x)}^{w}$, instead of $g(x)$, to the links connected to $x$.
The value of $w$ is chosen to speed up the convergence, 
see Ref.~\cite{Stack:2002ysv}
for the optimal value for $SU(3)$.

\subsection{The Weyl Symmetry of MAG}
\label{WeylSymm}

The Weyl group of $SU(N)$ is isomorphic to the permutation group acting on
the weights $\omega_i$ of the defining representation \cite{humphreys}. In
particular, for each $p$ in the symmetric group $S_N$, there exists 
$W_p \in SU(N)$ such that
\begin{equation}
W_p \, (\beta_i \cdot T)\, W_p^{-1} = \beta_{p(i)} \cdot T \;,
\end{equation}
where $p(i)$ denotes the image of the index $i$ under the permutation.
Our objective is to characterize the Cartan flux distributions associated
with the center-vortex network, together with their correlations. Since all
Cartan fluxes $2\pi \beta_i \cdot T$, $\beta_i = 2N \omega_i$, are physically
equivalent, i.e., they describe the same elementary $k=1$ center vortex, it is important for the
MAG step, which prepares the thermalized gauge configurations for monopole detection, to be Weyl-symmetric so as not to introduce any bias. 
In this regard, note that, using Eq.\,\eqref{weights}, our MAG functional
in Eq.\,\eqref{lattice mag} can be rewritten as
\begin{gather}
  {\cal E} =  \sum\limits_{x,\mu, q,p}\text{Tr }
  \big[U_\mu(x)\,T_q\,U^\dagger_\mu(x)\,T_p\big] \, 2N\sum\limits_{i}\omega_i|_q\,\omega_i|_p\\
   = 2N \sum\limits_{x,\mu,i}\text{Tr }
  \big[U_\mu(x)\, (\omega_i \cdot T)\,U^\dagger_\mu(x)\, (\omega_i \cdot T)\big]  \;,
\end{gather}
which puts in evidence the symmetry under permutations of the weights
$\omega_i$.

\section{Numerical Results and Perspectives}
\label{results}

In this section, we present our simulations and discuss our numerical
results, testing the prescriptions described in Section \ref{Cartanbased}
for the identification of monopoles.
In particular, we introduce a novel way to count monopoles, better suited
for comparing monopole-density results from the diagonal and Cartan-based
procedure, keeping track of their differences.
We also verify the Weyl symmetry (see Section \ref{WeylSymm}) of our results.

We performed our calculations for pure $SU(3)$ lattice gauge theory 
at several values of the inverse coupling $\beta$ and lattice extent
$L$, using the Wilson action
\begin{equation}
S\,=\,\frac{\beta}{3}\sum_{x,\,\mu<\nu}\text{Re Tr }
\Bigl[\mathbb{1} - U_\mu(x)\,U_\nu(x+\hat{\mu})\,
U_\mu^\dagger(x+\hat{\nu})\,U_\nu^\dagger(x)\Bigr]\,,
\end{equation}
where $U_\mu(x)\in SU(3)$ are the gauge-link variables,
$\mu,\nu$ are Lorentz indices and the lattice points $x$ are given in 
lattice units, i.e.\ $x_\mu=1,\dots,L\,$ for eah direction $\mu=1,\dots,4$.
Gauge-field updates were performed via the heat-bath algorithm according
to the Cabibbo-Marinari procedure, i.e.\ by using $SU(2)$ subgroups. 
Configurations were prepared discarding an intial
transient of 5000 thermalization sweeps, after which thermalized and
essentially independent configurations were generated. More specifically,
the decorrelation time length $\xi$ was estimated from the binning method
applied to the plaquette observable. Assuming a decorrelation window of
at least $2\,\xi$ time units for all our runs, we have taken measurements
separated by $200$ to $1000$ sweeps of the lattice. We note that this is
consistent with the intervals used in Refs.~\cite{athenodorou2021,lucini2001}. 

Lattice sizes ranged from $8^4$ to $20^4$, with $\beta$ values from 
$5.7$ to $6.2$. We generated 750 essentially independent configurations for
all runs, except for the $20^4$ lattice, for which 250 configurations were
produced. Statistical errors were computed using the binning and jack-knife
methods. In order to set the scale, we computed the string tension $\sigma$
from the static quark-antiquark potential. 
This was obtained from rectangular Wilson loops $W(R,T)$ of spatial and 
temporal lengths respectively $R$ and $T$, fitted to the expected asymptotic
form at large $T$
\begin{equation}
\log\, \bigl< W(R,T)\bigr> \,=\, -\,V(R)\,T+\text{const}\,.
\end{equation}
Loops with $R,T=1,2,\dots,L/2+2$ were considered, but only the ones 
with $T\ge 3$ were included in the fits, in order to reduce 
excited-state contamination. The string tension $\sigma$ was then
extracted by fitting the potential to the Cornell form
\begin{equation}
V(r)\,=\,V_0 \,+\,\sigma\,r\,+\,\frac{\gamma}{r},
\end{equation}
where $V_0$, $\sigma$, $\gamma$ are constants and $r$ is the
distance in physical units, given in terms of the lattice spacing $a$ 
as $r=R a$. For each coupling $\beta$ and lattice size $L$, the fitting
window was chosen so that the value of $a^2\sigma$ remained stable, 
while keeping $\chi^2/\text{d.o.f}\approx 1$. If needed, the physical scale
of the simulations may be estimated by equating the string tension to its
experimental value, of $(440\, \text{MeV})^2$.
In Table \ref{tab:table of the scale set} we show our run parameters and
the $\sigma$ values obtained from our fits. Except for the two coarser 
lattices (with $\beta=5.7, 5.8$), all our results correspond to an
effective lattice spacing $a\approx 0.1\,$fm and our values are in good
agreement with those of Ref.~\cite{lucini2001}.

\begin{table}[ht]
\centering
\begin{tabular}{|c|c|c|c|c|c|c|}
\hline
$\beta$ & $5.7$ & $5.8$ & $5.9$ & $5.93$ & $6.0$ & $6.2$ \\\hline
Lattice & $8^4$ & $10^4$ & $12^4$ & $16^4$ & $16^4$ & $20^4$\\\hline
Sweeps & 750 & 750 & 750 & 750 & 750 & 250\\\hline
$a\,\sqrt{\sigma}$ & 0.41(4) & 0.31(5) & 0.265(4) & 0.248(1) & 0.218(2) & 0.173(6)\\\hline
\end{tabular}
\caption{Our run parameters and corresponding measured values of the 
string tension, given in lattice units. Errors in parentheses correspond to
one standard deviation.}
\label{tab:table of the scale set}
\end{table}


We then fixed the MAG following the method described in Section \ref{MAG},
stopping the extremization process when the variation of the functional 
in Eq.\,\eqref{SU3funct} stayed below $10^{-9}$ for 50 consecutive iterations.
For comparison with Ref.~\cite{Stack:2002ysv}, we first implemented the
Abelian projection by the procedure described in Section \ref{Cartanbased}
(see Footnote \ref{symmetric_method}).
Accordingly, we have
initially performed the identification of monopoles following the 
{\em diagonal procedure}, explained in Section \ref{Cartanbased}.
In this case, the Abelian-projected link $C_\mu$ is parametrized by 
three dependent phases, given directly by its diagonal elements, as
in Eq.\,\eqref{abelianprojectedSU3}.
From the arguments of these phases, we evaluate the diagonal flux matrix
$F_{\mu\nu}$ and the associated quantity $\bar{F}_{\mu\nu}$, given by
shifting the phases to lie within the domain $[-\pi,\pi]^{\,3}$
(see Section \ref{compa}). The corresponding monopole current has components
\begin{equation}
j_\mu^{D}(\tilde{x})\vert_i\,=\,\frac{1}{4\pi}\,
\varepsilon_{\mu\nu\alpha\beta}\,
\partial_\nu\bar{F}_{\alpha\beta}(x)\vert_{ii}\;,
\label{jcurrentD}
\end{equation}
where $i=1,2,3$ labels the three charged currents --- the
quantities $j_\mu^{(i)}$ defined in Section \ref{Cartanbased} --- 
$\tilde{x}\in\tilde{\Lambda}$ represents a site on the dual lattice
(i.e.\ the center of a lattice hypercube) with ${\tilde{x}}_\mu=x_\mu+0.5$
for all $\mu$ and the superscript $D$ refers to the diagonal procedure. 
For each combination of $\tilde{x},\mu,i$, we say that the corresponding
cube around $\tilde{x}$ is charged if $j_\mu^{D}(\tilde{x})\vert_i$ is
nonzero, and include its absolute value in the computation of the quantity
$f_m$ defined in Ref.~\cite{Stack:2002ysv}. 
It counts the fraction of such charged cubes (which may be indexed by the
lattice links) carrying magnetic current, averaged over the three charge
components.
We measured the density $f_m$ for $\beta=5.9$ (lattice volume $12^4$) and
for $\beta=6.0$ (lattice volume $16^4$), respectively finding the values 
$1.26(1)\times 10^{-2}$ and $7.36(3)\times 10^{-3}$.
Both values are in good agreement with the ones in Table 4 of 
Ref.~\cite{Stack:2002ysv} within the errors. 

We also evaluate the {\bf density of monopoles}, defined as follows.
At a given site $\tilde{x}$, we fix $\mu=4$ and compute $j_4^{D}$ as
above, i.e.\ the total flux across the 3D lattice cube around $\tilde{x}$
defined by the remaining directions perpendicular to $\mu$.
We identify the presence of a monopole whenever there is at
least one nonzero component of $j_4^{D}(\tilde{x})$, defining the function
$\eta_{\mu}(\tilde{x})=1$ or $0$ if a monopole is detected
at $\tilde{x}$ or not.
More generally, by the rotation invariance of the lattice, one can also
average these monopole numbers over $\mu$.
The corresponding monopole density is then defined as the total count
divided by the total number of (dual) lattice points
\begin{equation}
\rho^D\,=\,\frac{1}{4V}\sum_{\tilde{x}\in\tilde{\Lambda}}
\sum\limits_{\mu=1}^4\, \eta^D_\mu(\tilde{x}) \;,
\label{rhocube}
\end{equation}
where $V=L^4$ denotes the four-dimensional lattice volume, and the superscript $D$ refers to the diagonal method.
Notice that Eq.\ \eqref{rhocube} simply counts how many cubes on the lattice
contain monopoles, while $f_m$ is affected by the internal index
structure of the current.
The above definition of the monopole density is better suited
for comparing results from the diagonal method and from our proposed 
Cartan-based method, since it is based only on the location of a monopole
and not on the nature of its charge.
Indeed, the diagonal method considers three (dependent)
charge components, while the Cartan-based method involves two components,
analyzed by projecting\footnote{
The function $\eta_{\mu}(\tilde{x})$ may be extended to identify, in the Cartan-based method, not only the presence of a monopole current at each point
$\tilde{x}$, but
also along which root direction it lies. This is done at the end of the
section.} along the roots $\alpha_{ij}$.
As argued in Section \ref{compa}, the two 
methods identify the same monopoles only if certain conditions are verified.
In particular, the diagonal method may overcount monopoles, and the
the density definition given in Eq.\ \eqref{rhocube} is suited for keeping
track of how many additional monopoles are measured by the diagonal method,
when compared to the same quantity obtained in the Cartan-based method.


\vskip 3mm
We can now proceed to implement the monopole detection following our proposed
prescription, the {\em Cartan-based procedure}, which takes into account the
Lie-algebra structure of the considered variables, as described before.
First, we perform the Abelian projection of the gauge links, in order to 
obtain the parameters $\theta_\mu(x)|_q$ ($q=1,2$) in 
Eq.\,\eqref{algebraicparametrization}, which in turn provide us with
the Cartan fluxes $f_{\mu\nu}$ of Eq.\,\eqref{algebraicplaquette},
given by
\begin{equation}
f_{\mu\nu}(x)=\theta_\mu(x)+\theta_\nu(x+\hat{\mu})-
\theta_\mu(x+\hat{\nu})-\theta_\nu(x).
\end{equation}
The gauge field $A_\mu(x)$ is related to the unprojected link
as $U_\mu(x)=e^{i a A_\mu}$. Its Cartan components are
$\mathcal{A}_\mu^q(x) = 2N\,\text{Tr}\left[A_\mu(x) T_q\right]$ and
the Abelian-projected link is defined as
\begin{equation}
C_\mu(x)\,=\,\exp\Bigl[\,i\,\mathcal{A}_\mu^q(x)\,T_q\,\Bigr]
\,=\,\exp\biggl[\,i\,\mathcal{A}_\mu^1(x)\,T_1
\,+\,i\,\mathcal{A}_\mu^2(x)\,T_2\,\biggr]\;,
\end{equation}
i.e.\ the two components of $\theta_\mu(x)$ in 
Eq.\,\eqref{algebraicparametrization} are identified as 
$\theta_\mu(x)\vert_q=\mathcal{A}_\mu^q(x)$. In practice, to a good
approximation, the angles $\theta_\mu(x)$ can be computed directly
from the diagonal elements of complete gauge link $U_\mu(x)$.

\begin{figure}[H]
\includegraphics[scale=.43]{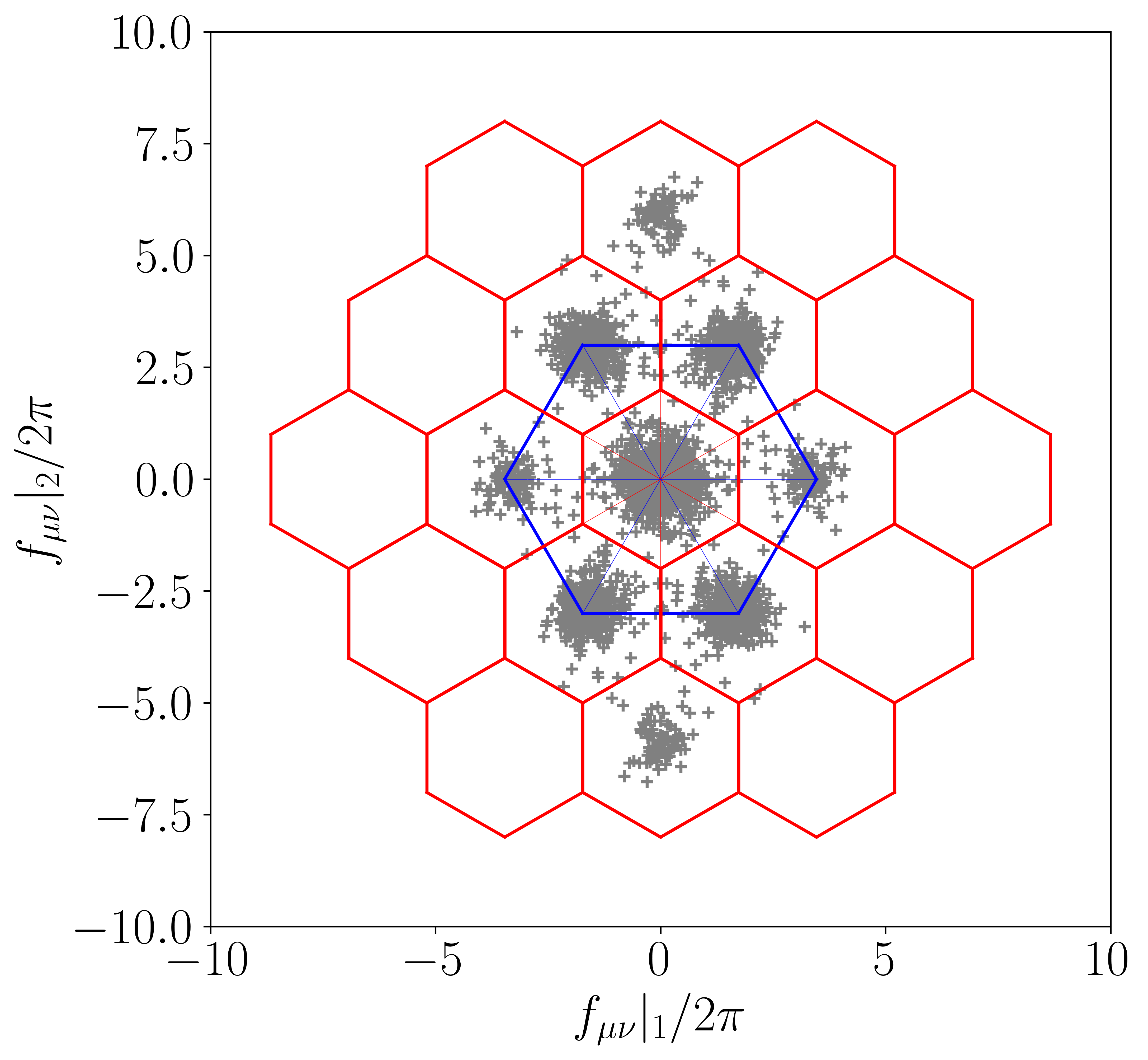}
\hspace*{2mm}
\includegraphics[scale=0.428]{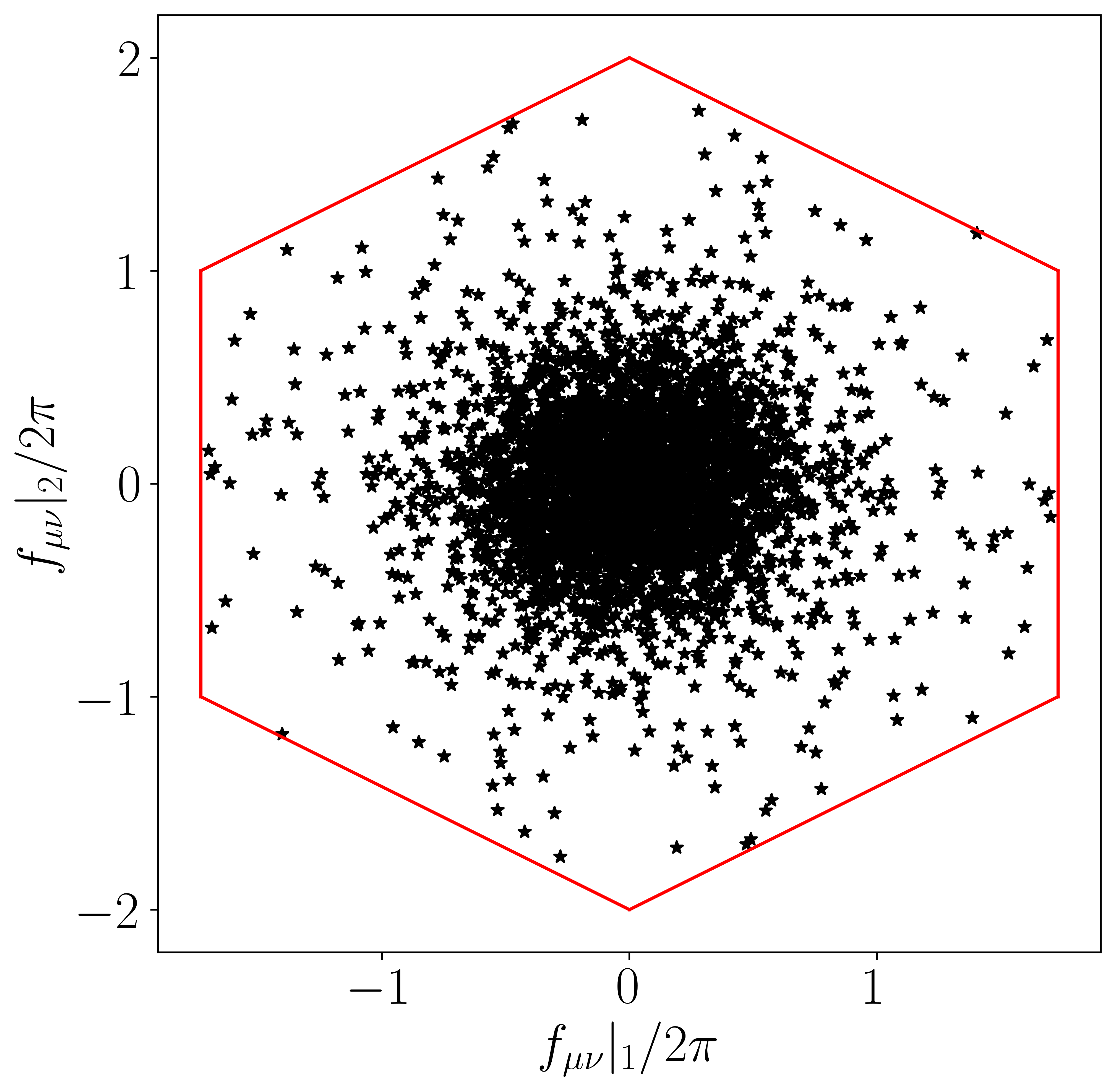}
\caption{Distribution of Cartan fluxes for a certain gauge configuration
before (left) and after (right) applying the Cartan-based DeGrand-Toussaint
algorithm. The plots show the two components of $f_{\mu\nu}/2\pi$ and
$\bar{f}_{\mu\nu}/2\pi$, respectively on the left and on the right.
See text for explanation.}
\label{hexagons}
\end{figure}

Next, as described in Section \ref{Cartanbased}, we must reduce the flux
tensor $f_{\mu\nu}$ to the fundamental cell, obtaining $\bar{f}_{\mu\nu}$
as in Eq.\,\eqref{HDecomposition2}. For convenience, let us consider
the normalized quantity $f_{\mu\nu}/2\pi$, whose two components may
be represented on a plane.
The fundamental cell is defined by reduced fluxes $\bar{f}_{\mu\nu}/2\pi$ 
having projections in the interval $[-1/\sqrt{3},1/\sqrt{3}]$
along the direction of the roots $\alpha_{ij}$, see Eq.\,\eqref{condi}.
The procedure is illustrated in Fig.\ \ref{hexagons}.
On the left, we show the distribution of the components of $f_{\mu\nu}/2\pi$
for all sites of a given configuration, with the fundamental hexagon at the
origin and its neighboring cells --- obtained by shifting it along the 
directions of the roots $\beta_{ij}=2N\alpha_{ij}$ by integer multiples 
of the interval $2/\sqrt{3}$ --- shown in red. 
We also show, in blue, the hexagon having $\beta_{ij}$ as its vertices.
As expected, the points are distributed around the center of the cells, with
a higher concentration in the fundamental cell and its nearest-neighbor
cells, which correspond to shifts by only one interval along $\beta_{ij}$.
Still, points located at second-neighbor cells are present,
albeit with much lower occurrence.
In order to bring the value of $f_{\mu\nu}/2\pi$ into the fundamental cell,
we sweep over the three roots $\alpha_{ij}$ and check the projection 
$f_{\mu\nu}\cdot\alpha_{ij}/2\pi$. If it is less than $1$ in magnitude,
the point is inside the fundamental cell. Otherwise, we subtract the
projection and follow the procedure until the reduced flux tensor
$\bar{f}_{\mu\nu}/2\pi$, shown in the figure on the right, is obtained.
Let us note that the points shown represent the flux values for
the plaquettes, while the monopole charge depends on the total flux
on the cube around a given point $\tilde{x}$ (at fixed $\mu$), yielding 
integer numbers.

Finally, we can use $\bar{f}_{\mu\nu}(x)$ obtained as above and the
analogue of Eq.\,\eqref{jcurrentD} to compute the monopole current for
the Cartan-based case $j_\mu^{C}=(j_\mu^{C}|_1,j_\mu^{C}|_2)$, given by
\begin{equation}
j_\mu^{C}({\tilde x})\,=\,\frac{1}{4\pi}\,
\varepsilon_{\mu\nu\alpha\beta}\,\partial_\nu\bar{f}_{\alpha\beta}(x) \;.
\label{eq:densidade1}
\end{equation}
Again, we identify the presence of a monopole whenever $j_\mu^{C}\ne (0,0)$
and define $\eta_{\mu}^{C}(\tilde{x})=1$ or 0 if a monopole is detected
inside the cube (along $\mu$) around $\tilde{x}$ or not, averaging over 
the directions $\mu$.
The corresponding monopole density is given by
\begin{equation}
\rho^C\,=\,\frac{1}{4V}\,\sum_{\tilde{x}\in\tilde{\Lambda}}
\sum\limits_{\mu=1}^4 \,\eta^C_\mu(\tilde{x}) \;.
\label{rhoCartan}
\end{equation}

\vskip 3mm
In Table \ref{densitiesCartcube} we
present our results for the densities $\rho^D$ and $\rho^C$, 
see Eqs.\,\eqref{rhocube} and \eqref{rhoCartan}, respectively computed
using the diagonal and the Cartan-based methods.
A sizeable difference between the densites is observed in the whole
range of $\beta$ and lattice sizes considered. 
In particular, the diagonal method always returns a significanlty
higher density (by 13\%-20\%) than the Cartan-based one. 
This is in agreement with our discussion of Section \ref{compa}, 
which shows how the diagonal method additionally counts monopoles
whose total fluxes may lie outside the Cartan subalgebra. 

\begin{table}[ht]
\centering
\begin{tabular}{||c|c|c|c|c|c|c|}
\hline
$\beta$ & Lattice & Sweeps & $\rho^D$ & $\rho^C$ & $|\rho^D-\rho^C|$\\\hline
5.7 & 8 & 750 & 0.0687(3) & 0.0607(3) & 0.0080(1) \\\hline
5.8 & 12 & 750 & 0.0379(1) & 0.0330(1) & 0.00496(4) \\\hline
5.9 & 12 & 750 & 0.02147(9) & 0.01836(8) & 0.00311(3) \\\hline
5.93 & 16 & 750 & 0.01805(5) & 0.01539(4) & 0.00266(1) \\\hline
6.0 & 16 & 750 & 0.01246(4) & 0.01055(4) & 0.00192(1) \\\hline
6.2 & 20 & 250 & 0.00450(4) & 0.00373(3) & 0.000764(9) \\\hline
\end{tabular}
\caption{Monopole densities $\rho^D$ and $\rho^C$, see
Eqs.\,\eqref{rhocube} and \eqref{rhoCartan}, calculated by counting 
monopoles and averaging over space directions and lattice volume $V=L^4$,
respectively for the diagonal and the Cartan-based methods.
Errors in parentheses correspond to one standard deviation.}
\label{densitiesCartcube}
\end{table}

\begin{figure}[ht]
\centering
\includegraphics[scale=.54]{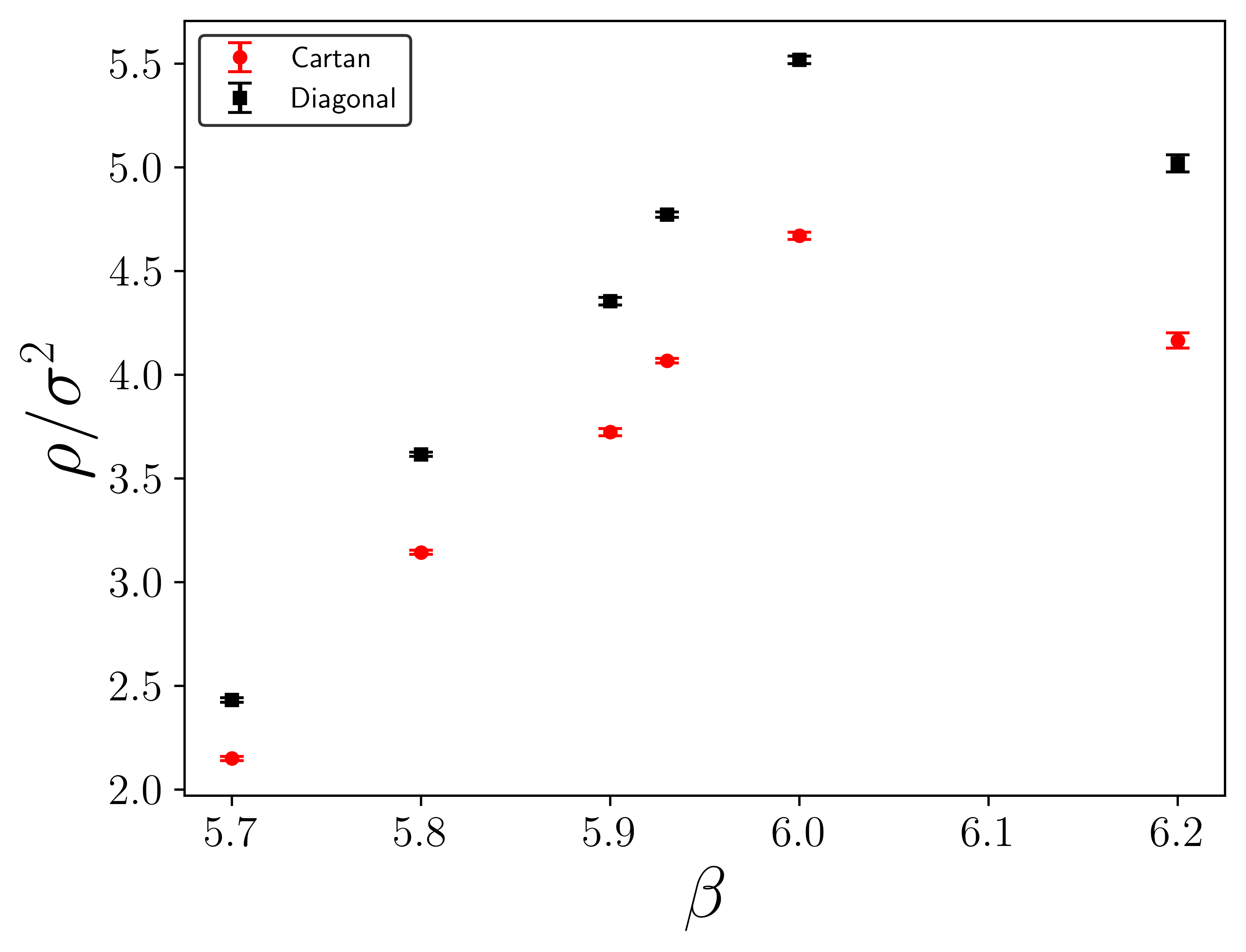}
\hskip 3mm
\includegraphics[scale=.54]{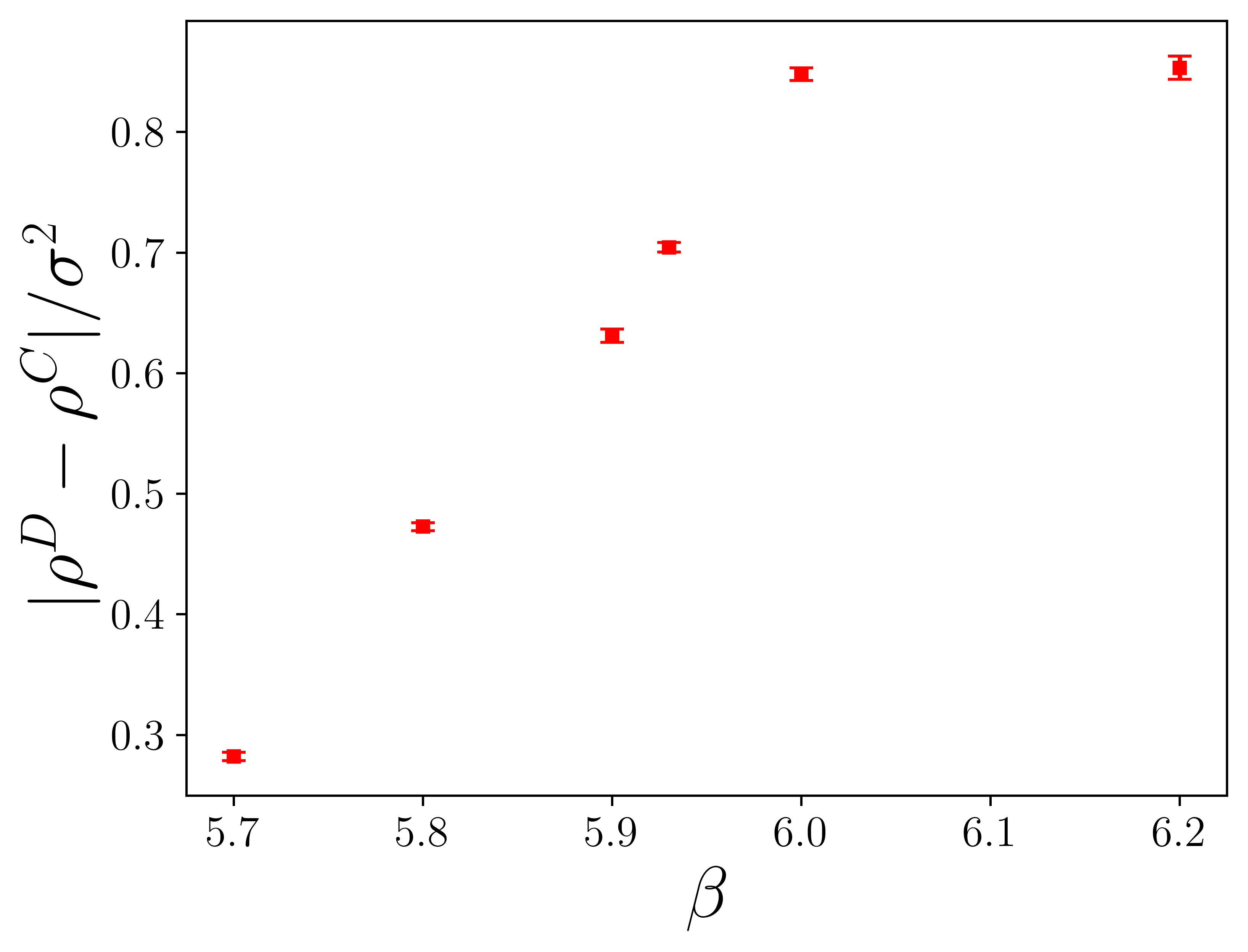}
\caption{Monopole densities $\rho^{C}$ and $\rho^{D}$ (see Table 
\ref{densitiesCartcube})
divided by the appropriate power of the string tension in order to 
absorb the lattice scale (left). The same is shown for the difference
$|\rho^{D} - \rho^{C}|$ (right).}
\label{comparison}
\end{figure}

Let us note that, as the continuum limit is approached, i.e.\ for increasing
values of $\beta$ or decreasing lattice spacing $a$, the difference 
$|\rho^D-\rho^C|$ seems to go to zero, perhaps suggesting that the
overcounting of unphysical monopoles by the diagonal method becomes
negligible and both methods detect the same magnetic defects. 
%
However, when we take the lattice scale into account, by computing each
value of the density in physical units, we see that this difference remains
finite.
In Fig.\ \ref{comparison} we plot the data from Table \ref{densitiesCartcube}
for the densities divided by the respective values of $a^4 \sigma^2$
(from Table \ref{tab:table of the scale set}), effectively considering
the density over the physical volume $(L a)^4$ of the lattice, i.e.\ 
converting to physical units.
Also, we see that the value of the density does not change appreciable
with $\beta$ as $a\to 0$, as shown in Fig.\ \ref{comparison}.
This is easily understood, since our runs for the larger lattices are in 
the scaling region and strong discretization effects are not expected for
observables related to low-energy (infrared) features of the theory.

As mentioned above, for the Cartan-based method, the distribution of
monopole charges can also be analyzed, by counting how many detected
monopoles lie along each root $\alpha_{ij}$. 
We do so for the projections along the three elementary charge directions
$\beta_{12}$, $\beta_{13}$ and $\beta_{23}$ (as well as for the corresponding
negative directions) and show our results in Table \ref{tab:occurrence}. 
We clearly see that Weyl symmetry is satisfied on average, since the values
for the different channels (positive or negative roots) agree within two
standard deviations. 


\begin{table}[ht]
\centering
\begin{tabular}{|cc||c|c|c|c|c|c|c|}
\hline	
$\beta$ & Lattice & $\beta_{12}$ & $\beta_{13}$ & $\beta_{23}$ & $-\beta_{12}$ & $-\beta_{13}$ & $-\beta_{23}$ \\\hline
5.7 & $8^4$ & 40.9(3) & 41.7(3) & 41.6(3) & 40.9(3) & 41.8(3) & 41.6(3) \\\hline
5.8 & $12^4$ & 113.4(5) & 114.2(5) & 113.8(5) & 113.7(5) & 113.9(5) & 114.1(5) \\\hline
5.9 & $12^4$ & 63.5(4) & 62.7(4) & 64.3(4) & 63.2(4) & 62.9(4) & 64.1(4) \\\hline
5.93 & $16^4$ & 167.2(7) & 168.6(7) & 168.3(7) & 167.7(7) & 168.0(7) & 168.8(7) \\\hline
6.0 & $16^4$ & 114.7(6) & 115.0(6) & 116.1(6) & 114.3(6) & 115.4(6) & 115.7(6) \\\hline
6.2 & $20^4$ & 100(1) & 99(1) & 99(1) & 101(1) & 99(1) & 100(1) \\\hline
\end{tabular}
\caption{Average number of monopoles per configuration carrying magnetic weight 
$\beta_{ij}$, obtained from the Cartan-based method.
Errors in parentheses correspond to one standard deviation.}
\label{tab:occurrence}
\end{table}

\section{Conclusions} 
\label{conclusions}

Over the years, the mechanism responsible for confinement has been extensively studied on the lattice. In this context, compelling evidence has accumulated in favor of center vortices, detected by center-projection methods, and monopoles, detected by Abelian projection. The vast majority of these studies have been carried out independently, focusing either on center vortices or on monopoles. The success of both scenarios, despite the observables not always being the same, has been attributed to a strong correlation between these two degrees of freedom. Indeed, lattice studies in $SU(2)$ have shown that Abelian-projected monopoles sit on center vortices, forming nonoriented center vortices or chains. Similar studies in $SU(3)$ found a correlation, although not as strong as in $SU(2)$.

On the theoretical side, a confinement mechanism based on a mixed ensemble was proposed, supported by an effective description of center-vortex worldsurfaces and monopole worldlines in 4D, as well as by a wave functional for Abelian-projected configurations. The ensemble is defined in terms of the simplest elementary center vortices and monopoles. In particular, the primary objects are center-charge $k=1$ vortices with a natural degeneracy: they are characterized by $N$ physically equivalent elementary Cartan fluxes, giving rise to a corresponding diversity of equivalent monopole charges. More precisely, elementary vortices carry the magnetic weights $\beta_i$, $i=1,\dots,N$, of the defining representation. This equivalence does not imply that the model can be formulated in terms of a single elementary Cartan flux, since correlations among vortices, between vortex worldsurfaces and monopole worldlines, and the fusion of monopole lines involve different $\beta_i$. As Weyl transformations permute these weights, it is impossible to align the corresponding fluxes along a single Cartan direction. The importance of Cartan-basis degrees of freedom was also observed in a recent lattice study extending condensed-matter ideas for gauge fixing \cite{Cucchieri:2025reg}, which showed that Cartan-projected group elements act as the non-Abelian analogue of plane waves.
\vskip 3mm

In this work, we presented a detection method well suited to exploring Cartan fluxes in Abelian-projected configurations. To this end, we first considered a parametrization of the links to extract the Cartan flux through a plaquette, fixing the Maximal Abelian gauge and defining the Abelian projection in a Weyl-symmetric manner. We then constructed a modified version of the DeGrand--Toussaint algorithm to detect monopoles on the lattice for $SU(N)$. Instead of processing the Abelian-projected plaquettes as if they were $N$ independent $U(1)$ fluxes, our method is based on the root lattice characterizing the Cartan subalgebra of $SU(N)$. In this way, we avoid spurious monopoles whose charges lie outside the Cartan sector. In short, our method exploits the natural periodicity of phases of elements of the Cartan subgroup, which lie on a scaled root lattice. This plays a role analogous to the $2\pi$ periodicity in compact QED and the $2\pi\sigma_3$ periodicity in $SU(2)$. Moreover, the analogue of the $[-\pi,+\pi]$ interval used to decompose and detect monopoles in compact QED is a unit cell given by a regular $(N-1)$-dimensional region whose vertices are characterized by the fluxes of elementary center vortices. For $SU(3)$, this region is a fundamental hexagon. Thus, at its core, the method reflects the possibility of detecting Cartan monopole charges together with their correlations with center vortices.

Numerical simulations performed with this method confirmed the Weyl symmetry among the elementary monopole charges, in the sense that none of them takes precedence over the others, and also revealed quantitative differences from the diagonal method. Both the locations and the total number of monopoles per lattice unit cube were found to differ, even as the continuum limit is approached. More specifically, the diagonal method consistently overcounts monopoles by approximately $13\%$ to $20\%$. Despite these differences, the two methods essentially agree whenever the detected monopoles carry fluxes in the Cartan subalgebra of $\mathfrak{su}(3)$. Consequently, observables involving monopole--center-vortex correlations, such as the fraction of monopoles located on center vortices, may differ appreciably depending on the detection method.

These developments establish a natural lattice framework for probing the mixed ensemble of oriented and nonoriented center-vortex worldsurfaces, which is likely to play a central role in the characterization of the $SU(3)$ Yang--Mills vacuum. They also open the way to systematic studies of monopole worldlines, their correlations with center-vortex worldsurfaces and their fusion properties, as well as the Cartan-flux distribution around the network of $Z(3)$ vortex guiding centers, which we are currently investigating \cite{inprep}. Verifying properties such as vortex–monopole correlation, flux collimation, and Weyl symmetry across oriented and non-oriented center-vortex ensembles will provide crucial constraints and guidance for identifying the relevant configurations responsible for confinement.

\section*{Acknowledgements}

The authors thank D. R. Junior and A. Cucchieri for helpful comments and
suggestions.
This work was partially supported by Fundação de Amparo à Pesquisa do Estado
de São Paulo (FAPESP), grants Nos.\ 2024/12768-5, 2023/11867-7 (GMS) and 
2024/19766-8 (RCST). Financial support from the Brazilian agency CNPq under 
Contracts Nos.\ 312736/2025-8 (TM) and 309971/2021-7 (LEO) is also gratefully
acknowledged.

\appendix

\section{The Cartan-Weyl basis of $\mathfrak{su}(N)$}
\label{Cartan-basis}

Given a Lie group $G$, one can define the vector space tangent to the group 
at the identity element. More concretely, when the Lie group is formed by 
matrices, consider a path such that 
\begin{equation}
    g(t) \in G\,,\qquad 
    g(0) = \mathbb{1}\;,
\end{equation}
where $t$ is a real parameter, and expand close to $t=0$, yielding
\begin{equation}
    g(t) \,=\, \mathbb{1} + X t + \dots\;.
\end{equation}
It can be shown that the set $\{ X \}$ generated by all possible paths forms
a real vector space that is closed under commutation, i.e.\ for $X$, $Y$
in the set, we have $[X, Y]$ also in the set. With this product, the vector space becomes a real Lie algebra $\mathfrak{g}$. A basis to expand $X$ can be introduced.
In particular, the product (commutator) of the basis elements is in the 
algebra, so it can be expanded in the Lie basis with real coefficients. 
It is always possible to define a representation of $G$ formed by maps 
${\rm Ad}_g$ that act on $\mathfrak{g}$, called the adjoint representation
of the group
\begin{equation}
    {\rm Ad}_g (Y) \,=\, g\, Y g^{-1}\,,\qquad
    Y \in \mathfrak{g}\;,\;g\in G\;.
\end{equation}
Moreover, if we consider a path $g(t)\in G$, the expansion of the adjoint representation of that path ${\rm Ad}_{g(t)}$ close to the identity
$\mathbb{1}$ leads to the adjoint representation of the Lie algebra
\begin{eqnarray}
    {\rm Ad}_{g(t)} (Y) & = &g(t)\, Y\, g(t)^{-1} \nonumber \\
    &=& (\mathbb{1} + X t + \dots)\, Y\, (\mathbb{1} -X t + \dots) \nonumber \\
    &=& Y + [X, Y]\, t
 + \dots \nonumber \\
    &=& (\mathbb{1} + {\rm Ad}_X\, t + \dots ) (Y)
    \;,
\end{eqnarray}
that is, the adjoint representation acts on $\mathfrak{g}$ according to
\begin{equation}
  {\rm Ad}_X(Y) \,=\, [X, Y] \;.  
\end{equation}

For $SU(N)$, $\{ X \}$  is the set of $N\times N$ anti-Hermitian matrices 
with vanishing trace. That is, the Lie algebra $\mathfrak{su}(N)$ has dimension 
$N^2-1$. The basis can be written as $\,iT_A$, 
$A = 1, \dots, N^2-1$, where $T_A$ are traceless Hermitian matrices
\begin{equation}
    X \,=\, i x_A \,T_A\;, \qquad x_A \in \mathbb{R}\;,
\end{equation}
where the index $A$ is summed over.
A particularly useful choice of basis, the so-called Cartan-Weyl basis, is 
defined as follows. Initially, find the largest commutative subalgebra, also
known as the Cartan subalgebra. For $\mathfrak{su}(N)$, its dimension is $N-1$.
For example, it can be generated by considering the diagonal basis elements, $q=1,\dots,N-1$,
\begin{equation}
  T_q\,=\, \frac{1}{\sqrt{2q\,(q+1)N}}\;\, {\rm Diag}(1,\dots,1,-q,0,\dots,0)\;,
  \label{dia}
\end{equation}
where the first $q$ elements are equal to $1$.

Before constructing the remaining elements of the Cartan-Weyl basis, it is 
useful to review some general concepts for any representation $D$
of $\mathfrak{g}$. By definition, a representation satisfies
\begin{align}
    D_{\left[X,Y\right]} \,=\,
    \big[D_X, D_Y\big] \;,
    \label{Drep}
\end{align}
so that the $T_q$'s are mutually commuting in any representation. Then, these maps 
admit a set of common eigenvectors  $|\lambda \rangle$, the so-called ``weight
vectors''. A weight is the tuple $\lambda$ formed by the eigenvalues 
corresponding to one common eigenvector\footnote{The weight vectors 
$|\lambda \rangle$ are similar to the spherical harmonics $|lm\rangle$, which
are common eigenvectors of the (commuting) angular-momentum operators $L^2$ 
and $L_z$, whose labels encode the corresponding eigenvalues
$l(l+1)\hbar^2,  m \hbar$.}
\begin{equation}
D_{T_q}|\lambda\rangle \,=\,\lambda|_q\, |\lambda\rangle\,,\qquad
\lambda \,=\, ({\lambda|}_1 , {\lambda|}_2 , \dots)\;. 
\end{equation}
For $\mathfrak{su}(N)$, the weights $\lambda$ of any representation $D$ are 
$N-1$ tuples.
Now, ${\rm Ad}$ is a representation, which means it has a set of weights. 
Of course, the Cartan basis elements $iT_p$ are common eigenvectors 
(with trivial eigenvalues):
\begin{align}
 {\rm Ad}(T_q)(T_p)= [T_q, T_p] = 0
     \;.
\end{align}
We shall denote the eigenvectors with nontrivial weights by $E_\alpha$, 
\begin{equation}
  {\rm Ad}(T_q)(E_\alpha) \,=\,  [T_q,E_\alpha]
   \,=\, {\alpha\vert}_q E_\alpha\,,\qquad \alpha \,=\, 
   ({\alpha|}_1, {\alpha|}_2, \dots)\;. 
    \label{rootsalpha}
\end{equation}
The nontrivial weights $\alpha$ of the adjoint representation are known as ``roots'', and $E_\alpha$ as ``root vector''. 
The roots and root vectors each come in pairs: for each $E_\alpha$, there is an $E_{-\alpha}$. It is convenient to define a weight as positive if the last nonvanishing component is positive. Later, we shall also use an ordering between weights: $\lambda' > \lambda $ if $\lambda' -\lambda $ is positive. Then, we can define independent Hermitian generators
\begin{equation}
    T_\alpha\,=\,\frac{E_\alpha+E_{-\alpha}}{\sqrt{2}}\;,
   \quad\;T_{\bar{\alpha}}\,=\,\frac{E_\alpha-E_{-\alpha}}{i\sqrt{2}}\;,
    \label{combi}
\end{equation}
where $\alpha$ is positive.
 They satisfy
\begin{equation}
    [T_q,T_\alpha]\,=\, i{\alpha|}_q\,T_{\bar{\alpha}}\,,\quad
    \big[T_q,T_{\bar{\alpha}}\big] \,=\, -i{\alpha|}_q\,T_{\alpha}\;.
\end{equation}
Together with the Cartan subalgebra, they form the Cartan-Weyl basis ${T_q,T_\alpha,T_{\bar{\alpha}}}$. The following triplets generate $\mathfrak{so}(3)$ subalgebras,
\begin{equation}
\frac{1}{\alpha^2}\, \alpha\cdot T\;,\qquad
\frac{1}{\sqrt{\alpha^2}}\, T_{\alpha}\;,\qquad
\frac{1}{\sqrt{\alpha^2}}\, T_{\bar{\alpha}}\;,
\label{subalg}
\end{equation}
where we used the notation $\alpha \cdot T \equiv {\alpha|}_q T_q$ and 
$\alpha^2 = {\alpha|}_q\, {\alpha|}_q $.
For $SU(N)$, there are ${N(N-1)}/{2}$ positive roots.

\section{Fundamental and Adjoint Weights of $\mathfrak{su}(N)$}
\label{App:weights}

When $D$ is the fundamental (defining) representation, there are $N$ different
weight vectors $|\omega_i \rangle$, $i=1, \dots, N$. When using the Cartan 
generators $T_q$ as diagonal matrices, each $|\omega_i \rangle$ can be written
as a column matrix: its elements are $1$ in the $i$th row and zero otherwise.
Then, there are $N$ defining weights
\begin{equation} 
\omega_i \,=\, ({T_1|}_{ii},\, {T_2|}_{ii},\dots,\, {T_{N-1}|}_{ii})  \;.
\label{wfun}
\end{equation}
As the generators are traceless, we have
\begin{equation} 
\omega_1+\dots +\omega_N\,=\,0\;.
\label{wfun2}
\end{equation}
Recalling Eq.\,\eqref{dia}, these weights are
ordered according to 
$\omega_1 > \omega_2 > \dots > \omega_N$ and satisfy (no sum over $i$)
\begin{equation}
\omega_i\cdot \omega_i \,=\, \frac{N-1}{2N^2}\,,\qquad
\omega_i \cdot \omega_j \,=\, -\frac{1}{2N^2}\,,\qquad i\neq j\;.
\label{fundwei}
\end{equation}
In particular, this leads to the useful property  
\begin{equation}
    e^{ i\!\: 2\pi\!\: 2N \!\: \omega_i \cdot T} 
\,=\, e^{- i\!\:\frac{2\pi}{N}}\, \mathbb{1}\;,
\end{equation}
which can be shown by computing the elements of the diagonal matrix
on the left-hand side, yielding
\begin{equation}
e^{i\!\; 2\pi\!\; 2N\!\; \omega_i \cdot T} |_{jj} \,=\,
e^{i\!\; 2\pi\!\; 2N\!\; \omega_i |_q \!\; T_q|_{jj}} \,=\,  
e^{i\!\; 2\pi\!\; 2N\!\; \omega_i \cdot \omega_j } \,=\, e^{- i\!\:\frac{2\pi}{N}} \;,
\end{equation}
where we used Eq.~\eqref{fundwei}, in the various cases
($j=i$ and $j\neq i$).

For the adjoint representation, there are $N(N-1)$ weight vectors with 
nontrivial weight (the root vectors $E_\alpha$). They are given by the 
$N\times N$ matrices that only have a nontrivial element at position $ij$, 
$i \neq j$, whose value is $1/\sqrt{2N}$. 
It is easy to see that the associated roots are
\begin{equation}
\alpha_{ij}\,=\,\omega_i -\omega_j
\label{difw}
\end{equation}
and satisfy $\alpha^2= \alpha \cdot \alpha=1/N$. The positive roots are 
characterized by $i < j$. 
The roots and fundamental weights satisfy (no sum over $i$)
\begin{equation}
2N\,\omega_i \cdot \alpha_{ij} \,=\,1\,,\qquad
2N\,\omega_i \cdot \alpha_{jk} = 0~~\;(j, k \neq i) \;,
\label{rela} 
\end{equation}
and the property
\begin{eqnarray}
\sum\limits_{i<j}\alpha_{ij}\vert_q\,\alpha_{ij}\vert_p &=& 
\frac{1}{2}\delta_{qp}\label{roots}\\[2mm]
\sum\limits_{i}\omega_i\vert_q\,\omega_i\vert_p &=& 
\frac{\delta_{qp}}{2N}\label{weights}
\end{eqnarray}

\subsection{$\mathfrak{su}(2)$}

The Cartan subalgebra of $\mathfrak{su}(2)$ is one-dimensional. 
It is spanned by the diagonal matrix
\begin{equation}
\label{T1}
    T_1 \,=\, \frac{1}{2\sqrt{2}} \left(\begin{array}{cc}
        1 & 0 \\
        0 & -1 
    \end{array}\right)\;,
\end{equation} 
cf.\ Eq.\,\eqref{dia}. Then, the weights of the defining representation 
have one component and are given by
\begin{equation}
\omega_1 \,=\, \frac{1}{2\sqrt{2}}\,,\qquad 
\omega_2 \,=\, -\frac{1}{2\sqrt{2}} \;.
\end{equation}
There is a positive root 
\begin{equation}
    \alpha_1 \,=\, \omega_1 -\omega_2 \,=\,  \frac{1}{\sqrt{2}} \;,
    \label{alphasu2}
\end{equation}
which corresponds to
\begin{eqnarray}
    E_{\alpha_1} &=& \frac{1}{2} \left(\begin{array}{cc}
        0 & 1 \\
        0 &0 
    \end{array}\right) \;.
\end{eqnarray}
Using Eq.\,\eqref{combi}, we get the relation with the usual Pauli matrices
\begin{equation}
T_1\,=\,\frac{\sigma_3}{2\sqrt{2}}\,,\qquad
T_{\alpha_1}\,=\,\frac{\sigma_1}{2\sqrt{2}}\,,\qquad
T_{\bar{\alpha}_1}\,=\,\frac{\sigma_2}{2\sqrt{2}}
\;. 
\label{mapa2}
\end{equation}

\subsection{$\mathfrak{su}(3)$}
\label{su3al}

The Cartan subalgebra of $\mathfrak{su}(3)$ is two-dimensional. It is 
spanned by the diagonal matrices
\begin{eqnarray}
T_1 &=& \frac{1}{2\sqrt{3}} \left(\begin{array}{ccc}
        1 & 0 &0\\
         0& -1 &0\\
         0 & 0 & 0
    \end{array}\right) \;,\\[3mm]
T_2 &=& \frac{1}{6} \left(\begin{array}{ccc}
        1 & 0 &0\\
         0& 1 &0\\
         0 & 0 & -2
    \end{array}\right)\;,
\end{eqnarray}
cf.\ Eq.\,\eqref{dia}. Then, the weights of the defining representation have 
two components and are given by
\begin{equation}
\omega_1 \,=\, \bigg(\,\frac{1}{2\sqrt{3}}\,,\;\frac{1}{6}\,\bigg)\;,\qquad
\omega_2 \,=\, \bigg(\!\!-\frac{1}{2\sqrt{3}}\,,\;\frac{1}{6}\,\bigg)\;,\qquad
\omega_3 \,=\, \bigg(0\,,\,-\frac{1}{3}\,\bigg) \;,
\label{w3}
\end{equation}
while the positive roots are
\begin{equation}
\alpha_1=\alpha_{13} \,=\, \bigg(\,\frac{1}{2\sqrt{3}},\;\frac{1}{2}\bigg)\;,\qquad
\alpha_2=\alpha_{23} \,=\, \bigg(\!-\frac{1}{2\sqrt{3}},\;\frac{1}{2}\bigg)\;,\qquad
\alpha_3=\alpha_{12} \,=\, \bigg(\frac{1}{\sqrt{3}},\,0\bigg) \;,
\label{roots3}
\end{equation}
whose ordered form is $\alpha_{1} > \alpha_{2} > \alpha_{3}$.
They correspond to
\begin{equation}
    E_{\alpha_1} \,=\,  \frac{1}{\sqrt{6}} \left(\begin{array}{ccc}
       0 & 0 &1\\
         0& 0 &0\\
         0 & 0 & 0
    \end{array}\right)\;,\qquad
    E_{\alpha_2} \,=\,  \frac{1}{\sqrt{6}} \left(\begin{array}{ccc}
       0 & 0 & 0\\
         0& 0 & 1\\
         0 & 0 & 0
    \end{array}\right)  \;,
\end{equation}
\begin{equation}
E_{\alpha_3} \,=\,  \frac{1}{\sqrt{6}} \left(\begin{array}{ccc}
       0 & 1 & 0\\
         0& 0 & 0\\
         0 & 0 & 0
    \end{array}\right)\;.
\vspace*{2mm}
\end{equation}
Using Eq.\,\eqref{combi}, we get the relation with the usual Gell-Mann 
matrices
\begin{equation}
T_1\,=\,\frac{\lambda_3}{2\sqrt{3}}\,,\qquad
T_2\,=\,\frac{\lambda_8}{2\sqrt{3}}\;,
\end{equation}
\begin{equation}
T_{\alpha_1}\,=\,\frac{\lambda_4}{2\sqrt{3}}\,,\qquad
T_{\alpha_2}\,=\,\frac{\lambda_6}{2\sqrt{3}}\,,\qquad
T_{\alpha_3}\,=\,\frac{\lambda_1}{2\sqrt{3}}\;,\nonumber
\vspace*{2mm}
\end{equation}
\begin{equation}
T_{\bar{\alpha}_1}\,=\,\frac{\lambda_5}{2\sqrt{3}}\,,\qquad
T_{\bar{\alpha}_2}\,=\,\frac{\lambda_7}{2\sqrt{3}}\,,\qquad
T_{\bar{\alpha}_3}\,=\,\frac{\lambda_2}{2\sqrt{3}}\;.
\end{equation}

\end{document}